\documentclass[longbibliography,nofootinbib,twocolumn,showpacs,amsmath,amstex,amssymb,mathfonts,superscriptaddress,prb]{revtex4-2}

\usepackage{babel,calc,amsmath,amsthm,amssymb,graphicx,subfigure,xcolor,comment}
\usepackage{mathdots}
\usepackage[T1]{fontenc}
\usepackage[unicode=true]{hyperref}
\usepackage{braket}
\usepackage{makecell}
\usepackage{booktabs}
\usepackage{mathtools}
\usepackage{textcomp}
\renewcommand{\arraystretch}{1.2}

\usepackage[normalem]{ulem}
\hypersetup{
	colorlinks=true,       		% false: boxed links; true: colored links
	linkcolor=blue,          	% color of internal links
	citecolor=blue,             % color of links to bibliography
	urlcolor=blue,          % color of external links 
}
\newtheorem*{theorem*}{Theorem}

\newtheorem*{corollary*}{Corollary}

\newtheorem*{lemma*}{Lemma}

\newtheorem*{proposition*}{Proposition}
\theoremstyle{definition}

\newtheorem*{definition*}{Definition}
\theoremstyle{remark}

\newtheorem*{remark*}{Remark}
\newif\ifdebug

\debugtrue

\ifdebug
\definecolor{zhliu}{rgb}{0.48, 0, 0.12}

\newcommand{\note}[1]{\textcolor{orange}{[#1]}}
\newcommand\delete{\bgroup\markoverwith{\textcolor{zhliu}{\rule[0.5ex]{2pt}{0.8pt}}}\ULon}

\else

\newcommand{\note}[1]{\ignorespaces}
\newcommand{\delete}[1]{\ignorespaces}

\fi

\begin{document}
	\renewcommand{\figurename}{Fig.}
	
	%\title{Photonic simulation of parafermionic Berry-phase statistics and contextuality}
	
	%\title{Parafermion contextuality and their statistics in a photonic platform}
	
	\title{Topological contextuality and anyonic statistics of photonic-encoded parafermions}
	
	%\title{Demonstrating topological contextuality with photonic parafermions}
	
	%-------------Contributing-------------
	\author{Zheng-Hao~Liu}
	\thanks{These two authors contributed equally to this work.}
	\author{Kai~Sun}
	\thanks{These two authors contributed equally to this work.}
	%email{ksun678@ustc.edu.cn}
	
	\affiliation{CAS Key Laboratory of Quantum Information, University of Science and Technology of China, Hefei 230026, People's Republic of China}
	\affiliation{CAS Centre For Excellence in Quantum Information and Quantum Physics, University of Science and Technology of China, Hefei 230026, People's Republic of China}
	
	\author{Jiannis~K.~Pachos}
	\email{j.k.pachos@leeds.ac.uk}
	\affiliation{School of Physics and Astronomy, University of Leeds, Leeds LS2 9JT, UK}
	
	%-------------Group-------------
	\author{Mu~Yang}
	\author{Yu~Meng}
	\author{Yu-Wei~Liao}
	\author{Qiang~Li}
	\author{Jun-Feng~Wang}
	
	%-------------Undergraduates-------------
	\author{Ze-Yu~Luo}
	\author{Yi-Fei~He}
	\author{Dong-Yu~Huang}
	\author{Guang-Rui~Ding}
	
	\affiliation{CAS Key Laboratory of Quantum Information, University of Science and Technology of China, Hefei 230026, People's Republic of China}
	\affiliation{CAS Centre For Excellence in Quantum Information and Quantum Physics, University of Science and Technology of China, Hefei 230026, People's Republic of China}
	
	%-------------Boss-------------
	
	%\author{Jing-Ling~Chen}
	%\email{chenjl@nankai.edu.cn}
	%\affiliation{Theoretical Physics Division, Chern Institute of Mathematics, Nankai University, Tianjin 300071, People's Republic of China}
	
	%\author{Xiao-Ye Xu}
	
	\author{Jin-Shi~Xu}
	\email{jsxu@ustc.edu.cn}
	
	\author{Yong-Jian~Han}
	\email{smhan@ustc.edu.cn}
	
	\author{Chuan-Feng~Li}
	\email{cfli@ustc.edu.cn}
	
	\author{Guang-Can~Guo}
	\affiliation{CAS Key Laboratory of Quantum Information, University of Science and Technology of China, Hefei 230026, People's Republic of China}
	\affiliation{CAS Centre For Excellence in Quantum Information and Quantum Physics, University of Science and Technology of China, Hefei 230026, People's Republic of China}
	
	\date{\today}
	
	\begin{abstract}
		
		Quasiparticle poisoning, expected to arise during the measurement of Majorana zero mode state, poses a fundamental problem towards the realization of Majorana-based quantum computation. Parafermions, a natural generalization of Majorana fermions, can encode topological qudits immune to quasiparticle poisoning. While parafermions are expected to emerge in superconducting fractional quantum Hall systems, they are not yet attainable with current technology. To bypass this problem, we employ a photonic quantum simulator to experimentally demonstrate the key components of parafermion-based universal quantum computation. Our contributions in this article are twofold. First, by manipulating the photonic states, we realize Clifford operator Berry phases that correspond to braiding statistics of parafermions. Second, we investigate the quantum contextuality in a topological system for the first time by demonstrating the contextuality of parafermion encoded qudit states. Importantly, we find that the topologically-encoded contextuality opens the way to magic state distillation, while both the contextuality and the braiding-induced Clifford gates are resilient against local noise. By introducing contextuality, our photonic quantum simulation provides the first step towards a physically robust methodology for realizing topological quantum computation. 
		
	\end{abstract}
	
	\date{\today}
	
	\maketitle
	
	\section{Introduction}
	
	Non-Abelian anyonic systems, in which information is non-locally encoded, offer an attractive approach to fault-tolerant quantum computation\,\cite{Preskill97, Kitaev03}. Anyonic quantum states are protected by topological order and are intrinsically immune to erroneous local perturbations\,\cite{Kitaev01}. Moreover, their non-Abelian statistics\,\cite{xgwen91} naturally induce topologically quantum gates in the encoded space that are immune to a wide range of control errors.
	Presently, Majorana zero modes remain the most conspicuous realization of the non-Abelian anyons. Several experimental signatures of Majorana fermions have been confirmed indicating their existence\,\cite{Sarma2015, Lutchyn2018}. Nevertheless, the realization of their non-Abelian statistics remains elusive. An important obstacle towards the experimental realization of Majorana braiding is that these systems suffer contamination by leaked quasiparticles from the environment, a mechanism known as quasiparticle poisoning\,\cite{Rainis12, Karzig2017, Lai2018, Lai2019, Menard2019}.
	This notorious mechanism dramatically reduces the coherence times of quantum states encoded in Majorana zero modes, thus creating significant challenges in the verification of their non-Abelian statistics by adiabatic braiding evolutions.
	
	%Due to the Ising anyonic fusion mechanism of the Majorana zero modes, all gates obtained by braiding of anyons are elements of the Clifford group, which can be efficiently simulated by classic computer according to the Gottesman-Knill theorem \cite{GK98}. As a result, the Majorana zero mode-based universal quantum computation can not be fully implemented under topological protection.
	
	To overcome the above demerit of Majorana fermions, the use of more exotic non-Abelian anyons is proposed. Parafermions\,\cite{Fendley12} with $\mathbb{Z}_n$ symmetry ($n$ positive integer), a natural extension of Majorana fermions that possess $\mathbb{Z}_2$ symmetry, provide symmetry-protected $n$-fold degenerate ground states which can be utilized to encode an $n$-level qudit. Remarkably, due to the existence of a composite topological charge for a pair of parafermions in superconducting fractional quantum Hall systems\,\cite{Tsui82, Laughlin83}, the quasiparticle poisoning in the parafermions-based quantum computation can be substantially suppressed\,\cite{Dua19}.
	Similar to Majorana zero modes, braiding of parafermions only generates the Clifford operators and is not solely competent for universal quantum computation. However, various approaches have been proposed to implement universal quantum computation with parafermions, for example, by obtaining Fibonacci anyon\,\cite{Stoudenmire15} from a nucleated phase of $\mathbb{Z}_3$ parafermions\,\cite{Nayak08}, or by asymptotically simulating the non-Clifford gates, such as the $\pi/8$-phase gate, with magic state distillation\,\cite{Bravyi05}. To determine if magic state distillation can be employed, one can use the onset of quantum contextuality\,\cite{KS67} that benchmarks resource suitability for universal quantum computation\,\cite{Howard14}. Moreover, the contextuality encoded in anyons is topologically protected, making the process robust to noisy input states.

	Realizations of parafermions have been proposed\,\cite{Clarke13, Mong14, Maghrebi15, Fleckenstein19, Laubscher20} by manipulating fractional quantum Hall states. 
	The realization of such a proposal is technically challenging, and only very recently the experimental signatures of generic anyons have been observed\,\cite{Bartolomei20, Nakamura20}.
	Quantum simulations\,\cite{Aspuru-Guzik12, Nori14}, on the other hand, have been very successful in demonstrating key ingredients of anyonic systems. Examples include the photonic implementation of anyonic braiding\,\cite{ksun16, Noh20} and quantum gate\,\cite{ksun18} by virtue of Majorana zero modes. In this article, we apply the idea of photonic quantum simulation, together with a novel form of dense encoding, to demonstrate the two key ingredients of the universal quantum computation with $\mathbb{Z}_3$ parafermions: the braiding to generate Clifford group and the contextuality in the topologically encoded space. The photonic simulator has the advantage of being virtually decoupled from the environment. Meanwhile, it supports the realization of accurate quantum gates, so it is ideal for implementing Berry phase evolutions that conclusively identify the anyonic statistics of parafermions. 
	When the parafermionic states are faithfully encoded with the beamsplitter network, their subsequent manipulations are robust against control errors due to their topological character.
	Moreover, our experiment comprises a quantum simulation of an intrinsically interacting system\,\cite{Pachos17}, the $\mathbb{Z}_3$ parafermions chain, compared to the Majorana chain, which can be described by non-interacting fermions\,\cite{Pachos18}. Due to the peculiarities of the photonic simulation, the encoding of the quantum states of the parafermion chain in an optical beamsplitter network lacks a tensor product structure, thus the size of the Hilbert space increases only polynomially with the employed photonic resources. Nevertheless, the recently empowered modulation\,\cite{Pierangeli19} and detection\,\cite{myang19} of hundreds of spatial modes paved the way to large-scale photonic quantum simulation. We envision that such photonic simulations will make it possible to implement prototype topological quantum algorithms in the near future\,\cite{ksun18}.
	
	\section{Braiding of Parafermion edge zero modes}
	
\begin{figure*}[t]
		\centering
		\includegraphics[width = 0.96\textwidth]{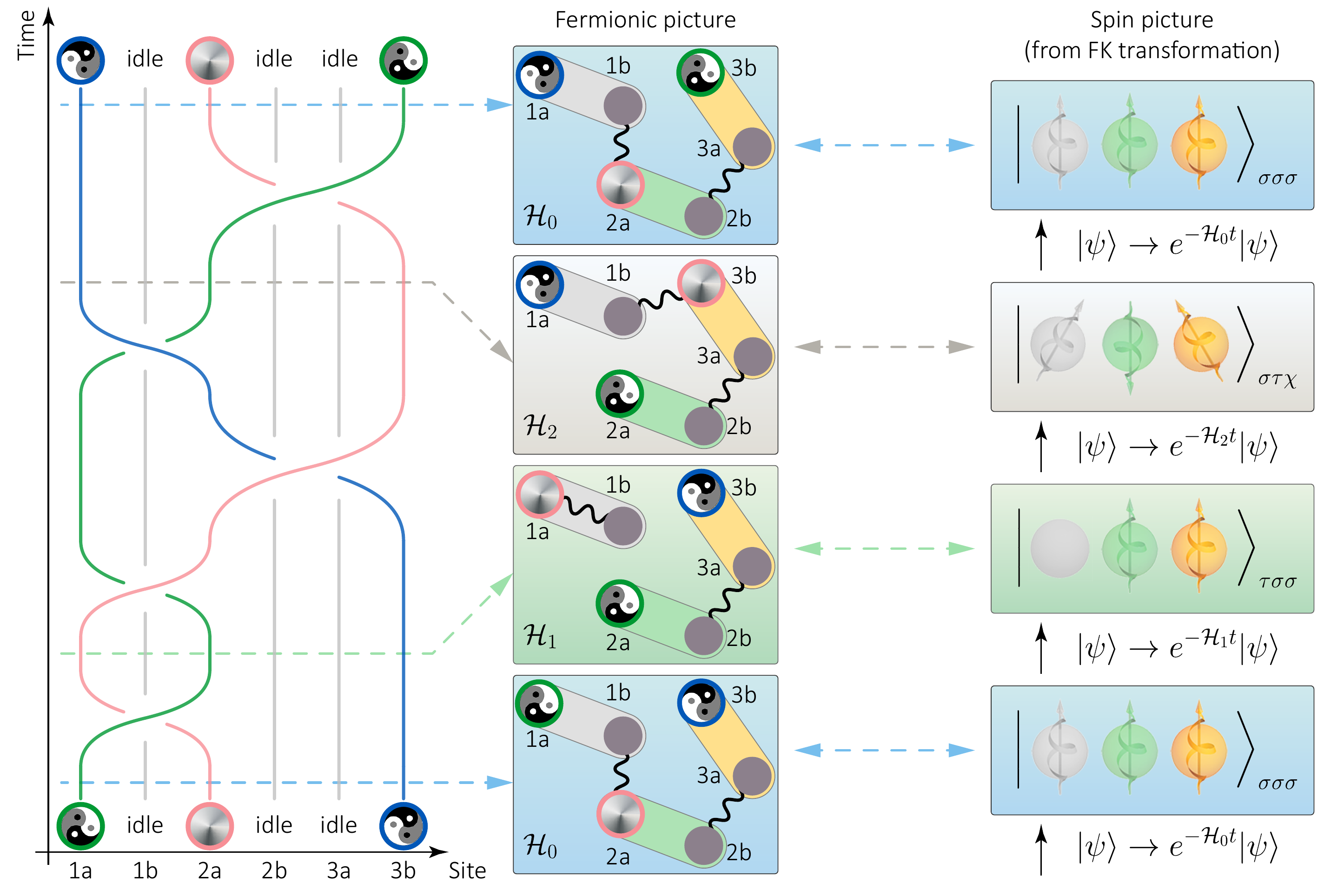}
		\caption{
			\textbf{Braiding of parafermion edge zero modes.} Left: braiding two parafermion edge zero modes located on a parafermionic chain's two remote sites using an auxiliary parafermionic site. The taichi symbols with green and blue edges denote two parafermion edge zero modes, the silver disk with pink edge denotes the auxiliary parafermion and the solid disks denotes idle parafermionic sites that do not participate in the braiding procedure. The braiding operation yields a Clifford gate on the encoded parafermions qutrit. Middle: physical realization of the braiding operation on an interacting parafermion chain. By adiabatically tuning the system Hamiltonian to $\mathcal{H}_0^\text{Pf}$, $\mathcal{H}_1^\text{Pf}$, $\mathcal{H}_2^\text{Pf}$ and again $\mathcal{H}_0^\text{Pf}$ (see Eqs. (\ref{eq:ham-parafermions-0}) and (\ref{eq:ham-parafermions-12})), the system evolves to different configurations that result in two parafermion edge zero modes being exchanged. Note that between $\mathcal{H}_1$ and $\mathcal{H}_2$ the green parafermion undergoes a type-2 Reidemeister move which does not affect the braiding process. The curled lines denote the interaction between different sites, and the rounded rectangles pair the parafermions that belong to a single site. Right: Applying the Fradkin--Kadanoff transformation on the chain yields a spin-1 quantum Potts model with $\mathbb{Z}_3$ symmetry. Its evolution can be effectively simulated by a series of imaginary-time evolutions which repeatedly project the system onto the ground state of new Hamiltonians. In each subpanel, the spin states corresponding to the empty sphere have an eigenvalue of 1 on the basis denoted as a subindex of the ket symbol, the spheres with upright arrows have the same eigenvalues, and the spheres with tilted arrows have spin eigenvalues that sum up to the sphere with inverted arrow (modulo 3; see the definition of the eigenstates after Eq. (\ref{eq:HamS})). 
The dashed horizontal lines link the configurations in the three different pictures at each stage of the evolution.
		}
		\label{fig:concept}
	\end{figure*}

	%The simulation of the braiding of two parafermions is demonstrated by first mapping the parafermion chain to the equivalent spin chain\,\cite{ksun16}. 
	Our starting model is a parafermion chain with two $\mathbb{Z}_3$ parafermions ($\alpha_{ja}$ and $\alpha_{jb}$) at each site $j$. The parafermion operators satisfy the commutation relation $\alpha_{ja}\alpha_{ka} = \omega \alpha_{ka}\alpha_{ja}$, $\alpha_{jb}\alpha_{kb} = \omega \alpha_{kb}\alpha_{jb}$ and $\alpha_{ja}\alpha_{kb} = \omega \alpha_{kb}\alpha_{ja}$ for $j<k$, where $\omega = e^{2\pi i/3}$.
	For simplicity and clarity, throughout this article we focus on a chain with three sites, $j=1,2,3$, and manipulate it in order to obtain the parafermion statistics as a Berry phase evolution. We start from the configuration subjecting to the following Hamiltonian
	\begin{align}
		\mathcal{H}_0^\text{Pf} = -e^{i\pi/6} (\alpha_{1b} \alpha_{2a}^\dagger + \alpha_{2b} \alpha_{3a}^\dagger) + \text{h.c.}.
		\label{eq:ham-parafermions-0}
	\end{align}
	This chain has two $\mathbb{Z}_3$ parafermion edge zero modes ($\alpha_{1a}$ and $\alpha_{3b}$) emerging at the edges of the chain, as illustrated in the middle panel of Fig.\,\ref{fig:concept}. The term ``zero'' mode is coined because these modes are free operators vanishing from the Hamiltonian, thus contributing zero energy shift when populated. The ground state of $\mathcal{H}_0^\text{Pf}$ is $3$-fold degenerate and can encode a non-local qutrit. For convenience, we choose the basis $\ket{\psi_l^\text{Pf}}$ ($l=0,1,2$) as the eigenstates of the parity operator $Q=\prod_k \alpha_{ka}^\dagger \alpha_{kb}$ with eigenvalues $\omega^l$.
	
	The non-trivial statistics of parafermions is manifested by the emergence of different phase factors between the above eigenstates when the parafermion edge zero modes, $\alpha_{1a}$ and $\alpha_{3b}$, are exchanged\,\cite{Hutter16}. To braid these two parafermion edge zero modes, we can adiabatically cycle between the Hamiltonians $\mathcal{H}_0$, $\mathcal{H}_1$ and $\mathcal{H}_2$, where
	\begin{align}
		\begin{alignedat}{9}
			\mathcal{H}_1^\text{Pf} &= -e^{i\pi/6} \alpha_{2b} \alpha_{3a}^\dagger -\alpha_{1a} \alpha_{1b}^\dagger + \text{h.c.}, \\
			\mathcal{H}_2^\text{Pf} &= -e^{i\pi/6} (\alpha_{2b} \alpha_{3a}^\dagger + \alpha_{3b} \alpha_{1b}^\dagger) + \text{h.c.}.
		\end{alignedat}
		\label{eq:ham-parafermions-12}
	\end{align}
	Here, the ferromagnetic term $\alpha_{1a} \alpha_{1b}^\dagger$ in (\ref{eq:ham-parafermions-12}) resembles the effect of a local external field, while the other terms are analogous to a paramagnetic interaction between neighbor sites. Therefore, all employed Hamiltonians have a clear physical interpretation with components that are amenable in quantum Hall systems. We also note that, in our configuration the external fields do not overlap with the interacting parts of the chain that can freely support parafermion edge zero modes. In the case they overlap the maximum tolerable external field is determined by the chirality of the interactions, represented by the phase factor of the $\alpha_{jb}^{\phantom{\dagger}} \alpha_{(j+1)a}^\dagger$ terms\,\cite{Fendley12}. Our choice of $\pi/6$ makes the parafermion edge zero modes most robust.
	
	We expound the effect of the adiabatic evolution, $\mathcal{H}_0^\text{Pf}\to\mathcal{H}_1^\text{Pf} \to\mathcal{H}_2^\text{Pf}\to\mathcal{H}_0^\text{Pf}$, on the parafermion edge zero modes, in the left panel of Fig.\,\ref{fig:concept}. First, by turning off the interaction between sites 1 and 2, we swap the parafermion edge zero mode initially on site 1a and the auxiliary parafermion at site 2a. Second, we interact the solitary site 1 with site 3, so the parafermion edge zero mode initially situated on site 3b is driven to site 1a which is now at the end of the interaction chain; finally, we transfer the parafermion edge zero mode from site 2a to site 3b by resetting the system's couplings to their configuration. Modulo the braiding with the auxiliary parafermion which has no effect, the evolution results in the exchange of the two parafermion edge zero modes $\alpha_{1a}$ and $\alpha_{3b}$.
	As the braiding operator and the parity operator can be simultaneously diagonalized\,\cite{Hutter16}, the braiding induces a Berry phase acting on the degenerate subspace spanned by $\ket{\psi_0^\text{Pf}}$, $\ket{\psi_1^\text{Pf}}$ and $\ket{\psi_2^\text{Pf}}$.

	\subsection{Simulation of parafermionic dynamics with a bosonic spin chain}
	
	Realization of parafermions based on the fractional quantum Hall system is highly non-trivial and still intractable with current technologies. Fortunately, through the Fradkin--Kadanoff transformation\,\cite{FK80}, the $\mathbb{Z}_3$ parafermions chain governed by the Hamiltonian $\mathcal{H}^\text{Pf}$ can be unitarily mapped to the quantum Potts model that describes a chiral spin-1 chain $\mathcal{H}^\text{S}$\, \cite{Fendley12}. As the Fradkin--Kadanoff transformation is non-local, it scrambles the topological order of the parafermions, so the corresponding spin-1 system is only symmetry protected, i.e., it is only naturally resilient against noise with $\mathbb{Z}_3$ symmetry. To reproduce adiabaticity during braiding we artificially constrain the dynamics of our photonic system into the ground state subspace of the spin chain by suppressing the wavefunction leaked into excited states. However, as the spectra of the parafermions and the spin-1 chains are identical, the statistical behavior, such as the Berry phase resulting from the braiding of parafermion edge zero modes, can be identically produced by the equivalent manipulations of the chiral spin-1 chain.
	
	The Fradkin--Kadanoff transformation maps the parafermionic Hamiltonians (\ref{eq:ham-parafermions-0}) and (\ref{eq:ham-parafermions-12}) into spin Hamiltonians with three sites of the form:
	\begin{align}
		\mathcal{H}_0^\text{S} &= -e^{i\pi/6} (\sigma_1 \sigma_2^\dagger + \sigma_2 \sigma_3^\dagger) + \text{h.c.}, \nonumber \\
		\mathcal{H}_1^\text{S} &= -e^{i\pi/6} \sigma_2 \sigma_3^\dagger -\tau_1 + \text{h.c.}, 
		\label{eq:HamS}
		\\
		\mathcal{H}_2^\text{S} &= -e^{i\pi/6} (\sigma_2 \sigma_3^\dagger + \sigma_1 \tau_2^\dagger \tau_3^\dagger \sigma_3^\dagger) + \text{h.c.}, \nonumber
	\end{align}
	\begin{align*}
		\text{with}~~~ \tau &=\begin{pmatrix}
			0 & 0 & 1 \\
			1 & 0 & 0 \\
			0 & 1 & 0
		\end{pmatrix},~~~ \sigma = \begin{pmatrix}
		1 & 0 & 0 \\
		0 & \omega & 0 \\
		0 & 0 & \omega^2
	\end{pmatrix},  \nonumber
\end{align*}
where the subindex corresponds to the three sites $j=1,2,3$ of the chain. As the Hamiltonians $\mathcal{H}^\text{S}$ and $\mathcal{H}^\text{Pf}$ have identical spectra, we know that the adiabatic evolution $\mathcal{H}_0^\text{S}\to\mathcal{H}_1^\text{S} \to\mathcal{H}_2^\text{S}\to\mathcal{H}_0^\text{S}$ of the spin system shall reproduce the Berry phase resulting from the adiabatic evolution of $\mathcal{H}^\text{Pf}$ that corresponds to the parafermion edge mode braiding. It is convenient to define three sets of bases for each site $j=1,2,3$ of the chain, namely $\ket{k}_\sigma$, $\ket{k}_\tau$ and $\ket{k}_\chi$, which are eigenstates of the operators $\sigma_j$, $\tau_j$ and $\sigma_j\tau_j$, respectively, with eigenvalues $\omega^k$. In terms of these states, the correspondence between the basis vectors of the parafermionic and spin picture in the degenerate spaces are as follows:
\begin{align}
	\begin{alignedat}{9}
		\ket{\psi_0^\text{Pf}} &\leftrightarrow \ket{\psi_0^\text{S}}=\frac{1}{\sqrt{3}}(\ket{000}_\sigma+\ket{111}_\sigma+\ket{222}_\sigma), \\
		\ket{\psi_1^\text{Pf}} &\leftrightarrow  \ket{\psi_1^\text{S}}=\frac{1}{\sqrt{3}}(\ket{000}_\sigma+\omega \ket{111}_\sigma+\omega^2 \ket{222}_\sigma), \\
		\ket{\psi_2^\text{Pf}} &\leftrightarrow \ket{\psi_2^\text{S}}=\frac{1}{\sqrt{3}}(\ket{000}_\sigma+\omega^2 \ket{111}_\sigma+\omega \ket{222}_\sigma).
		\label{eq:states}
	\end{alignedat}
\end{align}
The representation of the parafermions states in terms of spin states makes it possible to realize them with a photonic simulator, as we shall see in the following subsection. 

\subsection{Berry phase and dense encoding}
\label{sec:dense}

The first task of our quantum simulation is to obtain the Berry phase evolution that corresponds to the statistics of parafermions.
The Berry phases induced by braiding two parafermion edge zero modes are given in terms of projections by 
\begin{equation}
	\phi_{{\rm B}, l} = -\arg \braket{\psi_l^\text{S}|\Pi_1 \Pi_2|\psi_l^\text{S}},
	\label{eq:berry}
\end{equation}
where $l=0,1,2$ parametrizes the degenerate eigenstates of $\mathcal{H}_0^\text{S}$, while $\Pi_i$ projects the wavefunction on the ground states of the spin-1 Hamiltonian $\mathcal{H}_i^\text{S}$ ($i=1,2$) given in (\ref{eq:HamS}). Note that in general $\phi_{{\rm B}, l}$ is not the same for different $l$'s indicating the non-Abelian character of the parafermions. 
To realize this Berry phase, we implement the cyclic imaginary time evolution operator given by\,\cite{ksun16}
\begin{equation}
	e^{-\mathcal{H}_0^{\text{S}}t}\cdot e^{-\mathcal{H}_1^{\text{S}}t}\cdot e^{-\mathcal{H}_2^{\text{S}}t} \cdot e^{-\mathcal{H}_0^{\text{S}}t}.
	\label{eq:proj}
\end{equation}
The parameter $t$ is chosen to be large, so each term $e^{-\mathcal{H}^{\text{S}}t}$ acts as a projector $\Pi$ onto the space of ground states of $\mathcal{H}^{\text{S}}$; thus (\ref{eq:proj}) implements the Berry phase (\ref{eq:berry}). 

A vital characteristic of the imaginary-time evolution method is that we do not need to know the ground state of every involved Hamiltonian in order to carry out the simulation; this is because whenever two terms in Hamiltonian $\mathcal{H}^{\text{S}}$ commute, the imaginary time evolution operator (\ref{eq:proj}) can be further disassembled into two consecutive projections corresponding to each of the terms. Furthermore, as the system is finally projected back onto $\mathcal{H}_0^\text{S}$'s ground states, the first term of $\mathcal{H}_1^\text{S}$ and $\mathcal{H}_2^\text{S}$ in (\ref{eq:HamS}) can be completely omitted. As such, we only need to project the system three times onto the eigenstates of the operators $\sigma_j$, $\tau_j$ and $\sigma_j\tau_j$, which are $\ket{k}_\sigma \ket{k}_\sigma \ket{k}_\sigma$, $\ket{0}_\tau \ket{k}_\sigma \ket{k}_\sigma$ and $\ket{k}_\sigma \ket{l}_\tau \ket{l+k\, \text{mod}\, 3}_\chi $, respectively, with $k, l\in\{0, 1, 2\}$. This simulation procedure is shown in the right panel of Fig.\,\ref{fig:concept}.

The encoding of the parafermionic Berry phase in the imaginary-time evolution of a spin system can be further simplified before proceeding to the actual experiment. To elucidate this point, we note that although the set of vectors $\ket{k}_\sigma \ket{j}_\tau \ket{j+k}_\chi (j,k=0, 1, 2)$ lives in a nine-dimensional Hilbert space and it needs a qudit with $d=9$ to be encoded, projecting from this subspace onto the degenerate space of $\mathcal{H}_1^\text{S}$ or $\mathcal{H}_0^\text{S}$ only yields three independent coefficients. More concretely, the destination amplitude of a \textit{basis rotation} from the eigenstates of $\mathcal{H}_1^\text{S}$ or $\mathcal{H}_0^\text{S}$ to $\ket{k}_\sigma \ket{j}_\tau \ket{j+k}_\chi$ is solely determined by $k$, regardless of $j$, so the projection onto different $j$'s can be assimilated into the same physical mode, and only three orthogonal modes are required throughout the simulation. In other words, by densely encoding different eigenstates of $\mathcal{H}_2^\text{S}$, it is sufficient to simulate the braiding of a parafermionic chain through on-demand imaginary-time evolution on a physical qutrit ($d=3$). Our dense encoding method can be directly generalized to the quantum simulation of $\mathbb{Z}_n$ parafermions.

\section{A photonic Parafermion simulator}

The spin-1 chain system given in (\ref{eq:HamS}) can be directly encoded into an off-the-shelf photonic simulator. In particular, we encode the ground states of the homomorphic spin-1 chain into different spatial modes of a single photon, which constitutes a qutrit system. Since we only encode the information in the degenerated ground space, we can cast imaginary-time evolution ($e^{-\mathcal{H}^\text{S}t}$) corresponding to different Hamiltonians in the photonic setup to enforce the encoded spin system to their ground spaces. The consecutive implementation of photonic imaginary-time evolution is the core procedure of our dissipative quantum simulation. 

\begin{figure*}[tb]
	\centering
	\includegraphics[width = .88\textwidth]{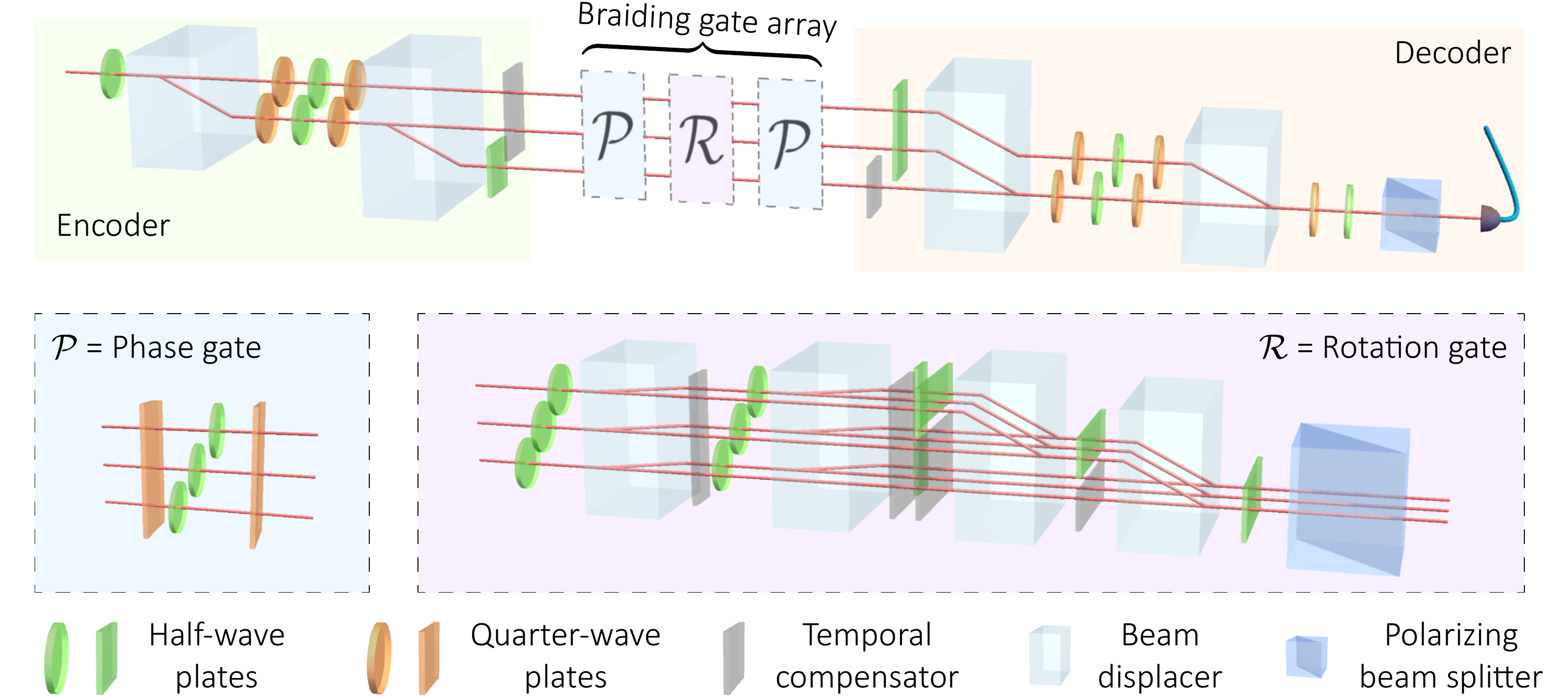}
	\caption{\textbf{Parafermion photonic simulator.} The qutrit state, $\ket{\psi^{\rm Pf}}$, of the parafermion edge zero modes, is encoded on photonic paths. The preparation of the qutrit state in the encoder is implemented using arrays of beam displacers. The half-wave plates and quarter-wave plates are exploited to control the complex amplitude of each mode. The photonic modes are then subjected to the required braiding evolution gates consisting of two phase gates $\mathcal{P}$ inducing phase unitaries between the three modes and a rotation gate $\mathcal{R}$ inducing a generic $\mathrm{SU}(3)$ rotation. Then, the architecture of beam displacers is again exploited in the decoder to recover the status of the parafermion edge zero modes by tomographic measurements. In the subplots with dashed boxes, we give the detailed photonic implementation of the $\mathcal{P}$ and $\mathcal{R}$ gates, where the individual modulation of the nine optical modes inside $\mathcal{R}$ is realized by additional amplitude filtering (not shown). Throughout the setup, we insert the temporal compensators to align the optical lengths of the modes so that the interference visibility is maximized and tilt them to introduce the necessary phase compensations. The brightness of the photon source in this experiment was about 100K counts per second on the maximum detection basis, and the integration time was 10 seconds per data point.}
	\label{fig:setup}
\end{figure*}

The optical simulator that realizes generic preparation, coherent evolution and the imaginary time evolution is shown in Fig.\,\ref{fig:setup}. The beam displacers, which are birefringent crystals separating the horizontally- and vertically-polarized propagating beams, are employed to prepare the photonic path qutrit. Before and between two sets of beam displacers, half-wave plates and quarter-wave plates are inserted for the on-demand control of the complex amplitude of the three modes. In principle, a $3^4$-dimensional Hilbert space is required to fully support the complete evolution of the $\mathbb{Z}_3$ parafermions braiding. By employing the method presented in\,\cite{ksun18}, which restricts to the ground state's dynamics, $3^3$ independent optical modes are required. We additionally employ the novel dense encoding presented in Sec.\,\ref{sec:dense} to further suppress the required resources in the simulator to only $3^2$ modes. This encoding extends the range of operations that can be simulated with our architecture and facilitates the realization of parafermions braiding evolutions.

The basic building blocks of the Berry phase evolution are the qutrit projectors $\Pi_i$, which implement the imaginary-time evolution. When the projector and the current Hamiltonian have the same set of eigenstates, then the projector is represented by a phase unitary realized by a non-dissipative phase gate $\mathcal{P}$, which induces relative phases between the three optical modes; a $\mathcal{P}$ gate is constructed using two quarter-wave plates with their optical axis fixed at $45^\circ$, sandwiching three adjustable half-wave plates acting on each path independently. When the projector does not share the same eigenstates with the current Hamiltonian, it must be implemented dissipatively. An arbitrary imaginary-time evolution is realized by a rotation gate $\mathcal{R}$ comprised of first a basis rotation and a subsequent dissipation of complex amplitudes of the eigenstates with higher energies. The $\mathcal{R}$ gate operates by first horizontally separating three vertically displaced optical modes with beam displacers twice in a row; the resulting nine complex amplitudes are then individually controlled; subsequently, these modes are vertically merged to obtain a horizontally arranged qutrit, with some photons dissipated as the result of imaginary-time evolution. This procedure is shown in the subpanels of Fig.\,\ref{fig:setup}.

Finally, in the decoding stage, another pair of beam displacers combines the three modes back. The additional half- and quarter-wave plates, together with a polarizing beam splitter, are employed for qutrit tomographic measurement, as shown in Fig. \ref{fig:setup}. As the measurement relies on interference between different optical modes and the photons used in our experiment have a coherence length of about 200\textmu m, the total optical length of each evolution channel must be aligned, so the differences between them are much smaller than the coherence length in order to preserve the temporal coherence and maximize interference visibility. In our experimental configuration, we introduce additional temporal compensators to compensate for the path differences introduced by the wave plates and optical misalignments and slightly tilt them to ensure the correct relative phase between different paths.

\subsection{Observation of parafermionic Berry phase in the simulator}

\begin{figure}[htbp]
	\centering
	\includegraphics[width = .48\textwidth]{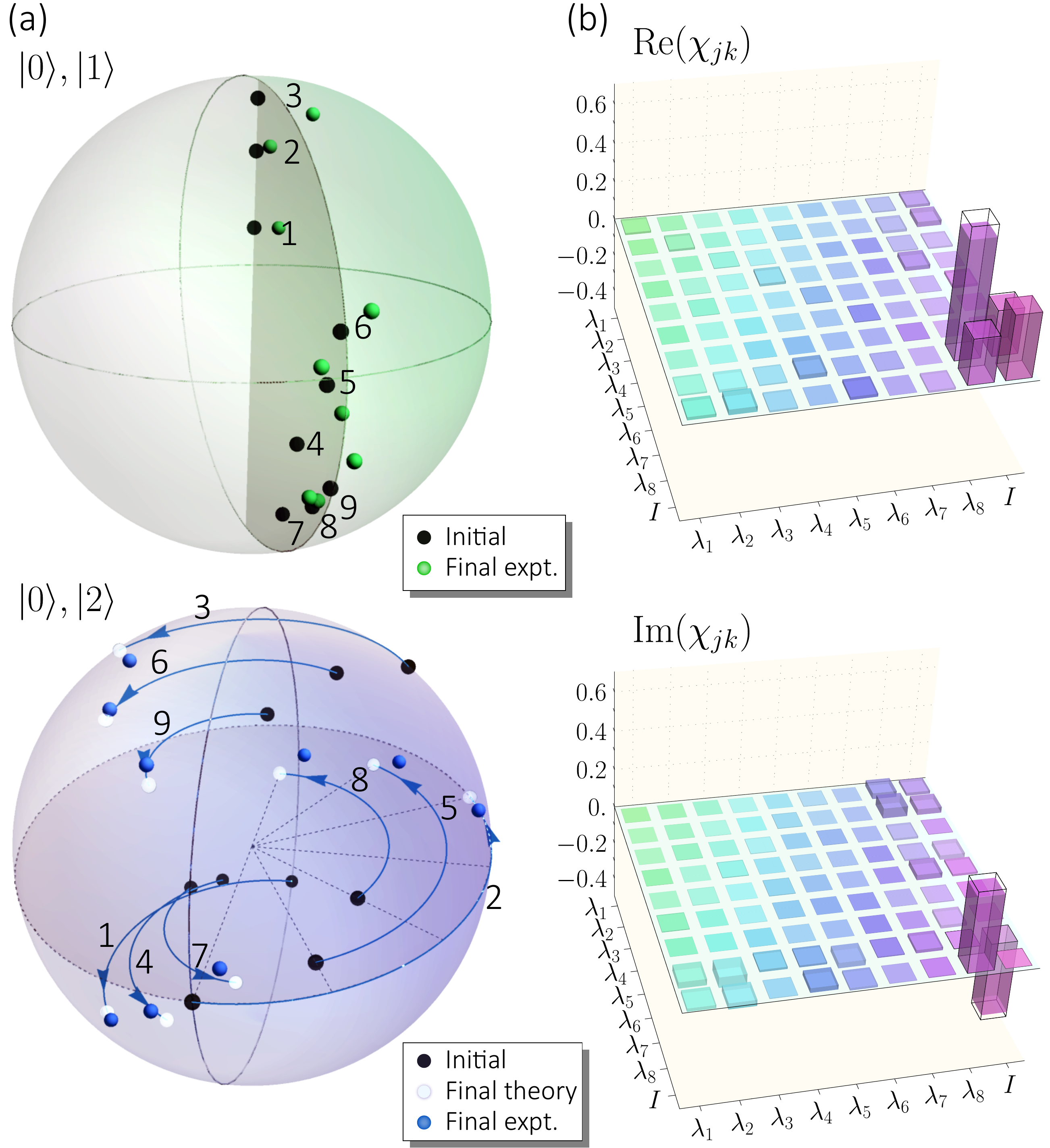}
	\caption{
		\textbf{Demonstration of two parafermion braiding}.
		(a) We employ nine states $\ket{\psi_j}$, $j=1, ..., 9$, to experimentally implement quantum state tomography before and after the braiding process. The top, green Bloch sphere shows the $\{\ket{\psi^\text{S}_0}, \ket{\psi^\text{S}_1}\}\otimes \{\bra{\psi^\text{S}_0}, \bra{\psi^\text{S}_1}\}$ parts of the density matrix, plotted in green points, while the bottom, blue Bloch sphere indicates the $\{\ket{\psi^\text{S}_0}, \ket{\psi^\text{S}_2}\}\otimes \{\bra{\psi^\text{S}_0}, \bra{\psi^\text{S}_2}\}$ parts.
		The black and white points denote the theoretical initial and final states, respectively. 
		In the top subplot, no significant relative phases can be seen in agreement with the theoretically predicted zero Berry phase between states $\ket{0}$ and $\ket{1}$ resulting from parafermion braiding.
		In the bottom subplot, each pair of points is connected by a circular arc of $2\pi/3$ central angle, which is the Berry phase induced between$\ket{0}$ and $\ket{2}$ by braiding the parafermions. The location of the points is calculated using the expectation values of the Gell-Mann matrices $\{\lambda_1,...,\lambda_8\}$ and the identity matrix, $I$.
		(b) The real (top) and imaginary (bottom) parts of the process matrix elements, $\chi_{jk}$, expressed in the Gell-Mann basis. The edge and filling of the cuboids show the theoretical values and experimental results from quantum process tomography, respectively.
	}
	\label{fig:braid}
\end{figure}

To investigate the effect of the parafermions braiding operations with the photonic simulator, we utilize nine generic qutrit states, $\ket{\psi_j}$, $j=1, ..., 9$ encoded in the three degenerate ground states of $\mathcal{H}_0^S$ and then implement the simulated braiding evolution on them. Quantum state tomography is then employed to identify the resulting states. The tomographic results, together with the theoretical predictions, are plotted in Fig.\,\ref{fig:braid}(a).
By translating the braiding evolution of the parafermions chain into three consecutive basis rotations and projections onto corresponding ground states, we find that two $\mathcal{P}$ gates and one $\mathcal{R}$ gate are required. From this procedure, we experimentally obtain that an extra phase of $\delta \phi_{{\rm B}, 2} = 2.061\pm0.128$ emerges between $\ket{\psi_2^\text{S}}$ and $\ket{\psi_0^\text{S}}$, while the phase between $\ket{\psi_1^\text{S}}$ and $\ket{\psi_0^\text{S}}$ is insignificant, taking the value $\delta\phi_{{\rm B}, 1} = 0.066\pm0.104$. These Berry phases correspond to $2\pi/3$ and 0, respectively, in agreement with the expected $\mathbb{Z}_3$ parafermion braiding matrix $\mathrm{diag}(1, 1, \omega) $\,\cite{Hutter16}, with an average fidelity of $94.9\%$. Note that $\phi_{{\rm B}, l}$ is not the same for different $l$'s emphasizing the non-Abelian character of the parafermions. The data in our experiment were recorded with single photon avalanche detectors; in order to estimate the counting errors, we assume a Poisson distribution for the counting statistics and resample the coincidence event numbers for 100 times to calculate the $1\sigma$ standard deviations of desired quantities.

Quantum process tomography is a reliable tool for the complete characterization of the braiding dynamics. From the tomography, we deduce the process matrix $\chi_{jk}$ of the evolution, spanned by the Gell-Mann basis $\{\lambda_1,...,\lambda_8\}$ and the identity ${I}$, with maximum likelihood estimation. The resulting process matrix corresponds to a relative phase of $2\pi/3$ between $\ket{\psi_2^\text{S}}$ and the other two states, as shown in Fig.\,\ref{fig:braid}(b). By comparing the result from the braiding quantum simulation with the expected theoretical value, we obtain the fidelity of the operation to be $93.4\%$.
Therefore, the braiding of the parafermions is faithfully simulated by a series of imaginary time evolution operators realized by mesoscale linear optical gates\,\cite{ksun16}.

We now comment on the sources of error present in the experimental simulation of parafermion braiding. The imaginary-time evolution of the system dissipates all optical modes that correspond to excited parafermionic states. As a result all the recorded errors occur inside the ground degenerate space, and are solely caused by experimental imperfections. From the quantum process tomography we see that the decoherence due to experimental imperfections plays a crucial role in the reduction of our measured fidelity. This arises in the photonic experiment from the different times the photons take to travel along different paths, causing the final interference fringes to have less than 100\% visibility. To a lesser degree, the decrease of fidelity is also caused by misalignment in the optical elements, element imperfections, and counting statistics.

\begin{figure*}[t]
	\centering
	\includegraphics[width = .96 \textwidth]{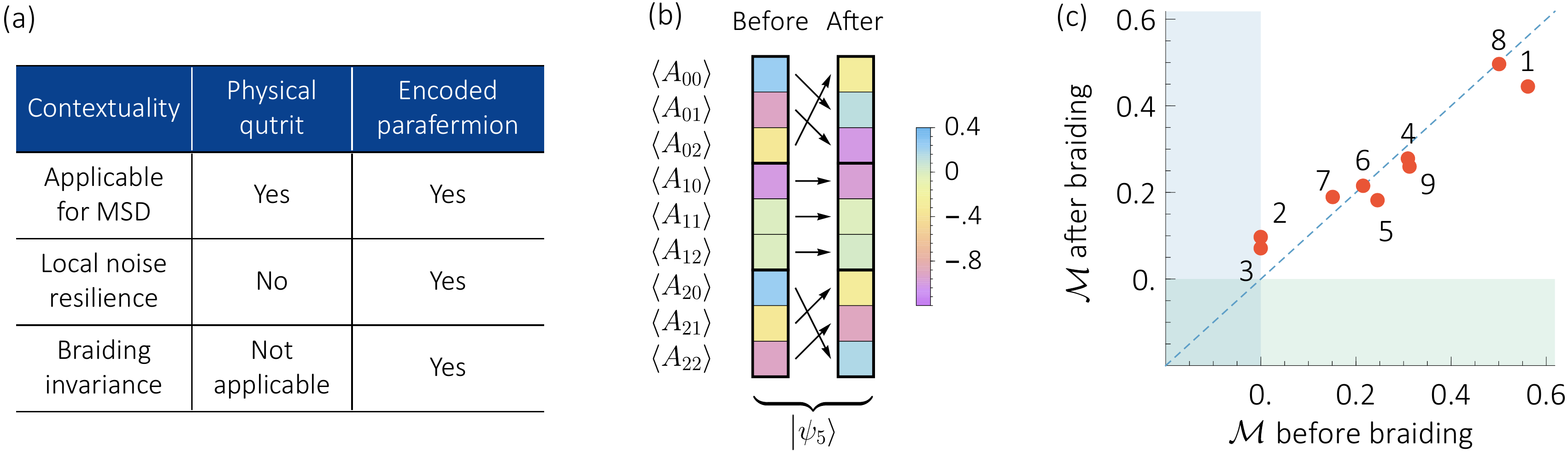}
	\caption{\textbf{Effect of braiding on contextuality.}
		(a) Comparison of contextuality encoded in non-topological (physical) qutrit and topologically-protected parafermionic qutrit. A physical qutrit subject to noise may lose its ability to manifest state-dependent contextuality and support magic state distillation (MSD). On the contrary, a qutrit encoded with parafermions is inherently immune to local noise. Applying Clifford operations on the parafermion-encoded qutrit preserves its property of contextuality, which is a crucial resource for quantum computation with magic state distillation. (b) The effect of braiding on the expectation values of the contextuality witness operators $A^{ij}, i,j\in\{0, 1, 2\}$ for parafermion edge zero mode states $\ket{\psi_5}$. The braiding induces a permutation between these values and thus preserves the contextuality witness $\mathcal{M}$.
		(c) The experimental results of the contextuality witness $\mathcal{M}$ for the nine sample states, $\ket{\psi_j}$, $j=1,...,9$, before and after braiding. The dashed line shows theoretical prediction. The error bars are too small to be displayed. The result indicates that braiding of $\mathbb{Z}_3$ parafermions preserves the state-dependent contextuality applicable for quantum computation with magic state distillation. The $1\sigma$ error bars deduced from a Poissonian counting statistics are too small to be visible in the graph.}
	\label{fig:resource}
\end{figure*}

\section{Parafermion Contextuality for quantum computation}

It is plausible that our experimental simulation of $\mathbb{Z}_3$ parafermions braiding can be extended to more complicated systems with several chains that encode topologically encoded qutrits in their parafermion edge zero modes just like the Majorana zero modes in \cite{ksun18}. In this case, braiding between arbitrary pairs of parafermion edge zero modes can generate all Clifford operators in the encoded space. As circuits with Clifford gates can be efficiently simulated by classical computers\,\cite{GK98}, exclusively parafermions braiding evolutions are not universal for quantum computation. A highly promising method for rectifying this problem is to employ magic state distillation on non-classical states generated e.g. by dynamical evolutions\,\cite{Sarma2015} or by employing half-fluxons\,\cite{Dua19}. Magic state distillation not only complements the Clifford circuit for producing universal quantum computation, but also has a fault-tolerant character. State-dependent quantum contextuality is a necessary criterion that identifies if magic state distillation is applicable in a circuit or not\,\cite{Howard14}. In other words, quantum contextuality and Clifford operations bootstrap the full power of universal quantum computation.

In this section, we experimentally demonstrate that a parafermionic universal quantum computation architecture is possible based on the fault-tolerant procedures of braiding and magic state distillation, as summarized in Fig.\,\ref{fig:resource}(a).  Assuming that we can generate quantum states that are not classically simulable, e.g., by activating interactions between parafermion edge zero modes, then magic state distillation can be applied at an arbitrary stage of the parafermions-based quantum computation to provide the non-Clifford gates required for universality. 

In the first subsection, we experimentally determine the contextuality properties of various topologically encoded states that enable magic state distillation. Specifically, we demonstrate that contextuality remains unaffected by the application of braiding-induced Clifford gates. In the second subsection, we elucidate that the parafermion encoded contextuality inherits the fault-tolerant feature from topological order. In the last subsection, we use another contextuality criterion to self-test the encoded state, which best describes its merit of local noise resilience. Therefore, we demonstrate that fault-tolerance extends to all the necessary elements of our parafermionic quantum computation architecture.

\subsection{Contextuality and its interplay with braiding}

Quantum contextuality\,\cite{KS67} is one of the most profound non-classical features of quantum theory. It states that an observable cannot have a predefined value regardless of the specific orthonormal basis used to measure it. The discrepancy between quantum and classical theories, where the observables have contextual-independent values, is revealed by the violation of non-contextual hidden-variable (NCHV) inequalities\,\cite{Budroni21}. Given a set of projective measurements, a NCHV inequality can always be formulated in the light of a recent graph-theoretic approach\,\cite{CSW14}. In particular, when the set of projectors corresponds to the qudit stabilizing group, the resulting NCHV inequalities bound the set of classically simulable states. As a result, a state that violates the inequality is suitable for quantum computation with magic state distillation\,\cite{Howard14}.

The construction of this NCHV inequality relies on the stabilizer formalism. Explicitly, the Weyl-Heisenberg displacement operators for a qutrit system are defined as
\begin{align}
	D_{x, z}=\omega^{2^{-1}xz}\tau^x\sigma^z,~~~ \{x, z\}\in\{0, 1, 2\}.
\end{align}
Each of these operators has a spectrum of $\{1, \omega, \omega^2\}$. For convenience, we denote the list of displacement operators $\mathbf{D}=\{D_{0, 1}, D_{1, 0}, D_{1, 1}, D_{1, 2}\}$, whose eigenstates span a complete set of mutually unbiased bases. By exploiting the above notation, a witness of contextuality can be constructed from the operators
\begin{equation}
	A^\mathbf{r}=\openone_3-\sum_{j=1}^4\Pi_j^{r_j}, 
\end{equation}
where $\Pi_j^{r_j}$ is the projector of the eigenstate $\omega^{r_j}$ of the $j$-th element in $\mathbf{D}$. The vector $\mathbf{r}$ is in turn defined as $\mathbf{r}=x\mathbf{a}+z\mathbf{b}$ with $\mathbf{a}=\{1, 0, 1, 2\}, \mathbf{b}=-\{0, 1, 1, 1\}$ and $\{x, z\}\in\{0, 1, 2\}$. For simplicity, we also adopt the shorthand notation $A^{xz}=A^{x\mathbf{a}+z\mathbf{b}}$. There are in total 9 witness operators, and the one with the maximum expectation value determines the suitability of a quantum state for magic state distillation. In particular, the NCHV inequality that identifies the magic state distillation resource states reads
\begin{align}
	\mathcal{M} &\equiv \max_{\bf r} \mathrm{Tr}\left[A^\mathbf{r} \rho \right] \overset{\mathrm{NCHV}}{\leq} 0,
	\label{eq:resource}
\end{align}
where $\rho$ is the quantum state to be considered. Because the projectors $\Pi_j^{r_j}$ of stabilizing operators form a closed group under Clifford operations, the witnesses $\mathcal{M}$ of contextuality are also unaffected by Clifford gates applied on the state $\rho$.

Experimentally, we apply the contextuality test~(\ref{eq:resource}) to the nine photonic states $\rho_i =\ket{\psi_i}\!\!\bra{\psi_i}$, $i=1,...,9$ shown in Fig.~\ref{fig:braid}(a), and determine the value of $\mathcal{M}$ for these states before and after braiding with the quantum simulator in order to determine their contextuality properties.
Fig.\,\ref{fig:resource}(b) illustrates the dynamics of the contextuality witness operators $A^{ij}$, with $i,j=\{0,1,2\}$ during braiding, which form a closed subgroup, acquiring each another's values. Direct comparison between these two groups of results shows that the values of the contextuality witnesses $\mathcal{M}$ are almost not affected by the process of braiding, as shown in Fig.\,\ref{fig:resource}(c). Consequently, the Clifford operations induced by braiding parafermions do not affect the state-dependent contextuality required for magic state distillation and can be safely performed at all stages of the computation.

\subsection{Observation of fault-tolerant contextuality}

The omnipresent local noise in a non-topological quantum system may diminish the observed contextuality, necessitating more copies  of resource states to achieve a target fidelity in magic state distilation\,\cite{Bravyi05}, or even completely hinder its implementation when contextuality is not present. In contrast, quantum states encoded in parafermionic systems are topologically protected. It is thus expected that the contextuality properties of these states are topologically immune enabling the application of magic state distillation even in the presence of local noise. 

To demonstrate this aptitude of parafermion systems, we apply local noise to both a topologically-encoded and a non-topological qutrit, and compare their response to noise. In the topological system, the flip of a single parafermion is prevented by parity conservation, so only simultaneous flips of two sites are allowed. As a result, the parafermionic system is susceptible to two possible forms of local noise: local hopping errors, which occur simultaneously on two adjacent sites, and local phase errors, which act on single sites. In order to characterize their behavior, it is convenient to define the Fock parafermion operators, $C_j$, associated with the canonical parafermions operators $\alpha_{ja}$ and $\alpha_{jb}$ by 
\begin{equation}
	C_k = \dfrac{2}{3}\alpha_{ka}-\dfrac{1}{3} \displaystyle\sum_{m=1}^{2} \omega^{m(m+1)/2} \alpha_{ka}^{m+1} {\alpha_{kb}^{\dagger}}^m.
\end{equation}
Then, the local hopping errors and the local phase errors can be expressed as $C_k^\dagger C_{k+1}$ and, $C_k^\dagger C_k$, respectively\,\cite{Cobanera14}, occurring with a certain probability. 

We first consider analytically the effect of hopping noise, $C_1^\dagger C_2$, between sites 1 and 2. Through the Fradkin--Kadanoff transformation, the corresponding operator in the spin picture reads 
\begin{equation}
	X^\text{S} = \frac{1}{9}\sigma_1^\dagger(2-\tau_1-\tau_1^\dagger)\tau_1 \sigma_2(2-\tau_2-\tau_2^\dagger),
\end{equation}
where the superscript S indicates that the operator is expressed in the spin picture.
If we apply the hopping error with probability $p$ 
\begin{equation}
	\rho \to X^\text{S}(\rho)=(1-p)\rho + p X^\text{S}\rho {X^\text{S}}^\dagger 
\end{equation}
on the three computational basis states we find:
\begin{align}
	\bra{\psi_i^\text{S}} {X}^\text{S}(\ket{\psi_j^\text{S}}\!\!\bra{\psi_j^\text{S}}) \ket{\psi_i^\text{S}} &= (1-p)\delta_{ij},
	\label{eq:prob}
\end{align}
where $\delta_{ij}$ is the Kronecker-$\delta$ function. Hence, applying the neighborhood hopping operator only causes the parafermions to leak out of the degenerate subspace and has no effect on the topological qutrit up to a state re-normalization. Similarly, the local phase noise on the parafermionic system with probability $q$ takes the form
\begin{equation}
	\rho \to {Z}^\text{S}(\rho)=(1-q)\rho + q Z^\text{S}\rho {Z^\text{S}}^\dagger, 
\end{equation}
with
\begin{equation}
	Z^\text{S} = \frac{1}{3} (3\openone-\tau_1-\tau_1^\dagger).
\end{equation}
Direct calculation shows that the result in (\ref{eq:prob}) still holds; namely, the dephasing noise always gives rise to an increase in system's energy and is thus corrected during imaginary time evolution.

To understand the protection exhibited by the topologically encoded qutrit, we compare it to a normal qutrit state subject to noise. We act on this state with two error models that resemble the errors in the parafermionic system, namely, the probabilistic flip error caused by the shift operator, $\tau$, occurring with probability $p$, and dephasing errors caused by the clock operator, $\sigma$, occurring with probability $q$, where $\tau$ and $\sigma$ are given in (\ref{eq:HamS}). 
Their corresponding effects on a state $\rho$ can be described by the superoperators 
\begin{equation}
	\mathrm{T}(\rho) = (1-p)\rho + \frac{p}{2}(\tau \rho \tau^\dagger + \tau^\dagger \rho \tau),
\end{equation}
and 
\begin{equation}
	\Sigma(\rho) = (1-q)\rho + \frac{q}{2}(\sigma \rho \sigma^\dagger + \sigma^\dagger \rho \sigma).
\end{equation}
While the qutrit error operators $\tau$ and $\sigma$ are in principle different to the operators $X$ and $Z$ acting on the parafermion qutrits, they correspond to the type of errors expected to emerge in typical physical realizations.

\begin{figure}[tbp]
\centering
\includegraphics[width = .99\columnwidth]{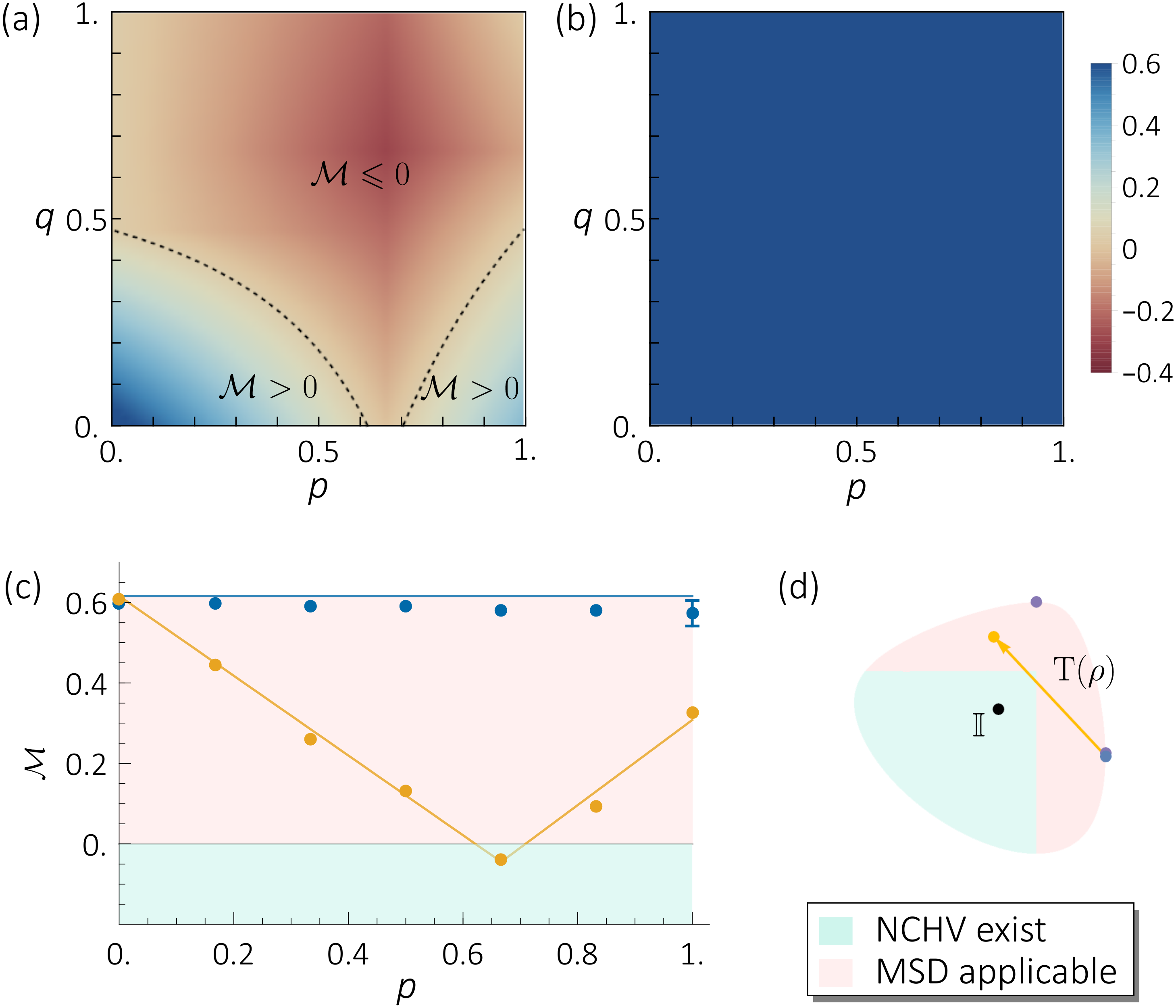}
\caption{
\textbf{Topologically-protected contextuality.}
Response of contextuality witness ${\cal M}$ to generic noise. (a) Behavior of ${\cal M}$ in the presence of composite noise $\mathrm{T}\circ\Sigma$ acting on non-topological qutrit. The dashed curves mark the classical bound of the NCHV inequality $\mathcal{M}\leqslant0$ that distinguishes quantum states useful in magic state distillation.
(b) Composite noise ${X}^\text{S}\circ {Z}^\text{S}$ acts on a parafermionic qutrit, with ${\cal M}$ exhibiting stability with respect to any amount of noise. (c) Experimental result of the contextual witness $\mathcal{M}$ to the amount of hopping (flip) noise. The blue (yellow) line and points correspond to the theoretical predictions and experimental data, respectively, in the parafermionic (non-topological) system. The results here correspond to the $q=0$ lines in (a) and (b).
(d) visualization of the effect of noise in the qutrit state space. The two coordinates correspond to two witnesses of contextuality in (\ref{eq:resource}). The purple points denote two of the magic states, the point marked $\mathbb{I}$ denotes the maximally mixed state, and the orange (blue) point represent the final non-topological (parafermionic) state at $p\to1$. The pink (cyan) area corresponds to the states which are (not) applicable for magic state distillation. The trajectory of the non-topological qutrit under flip noise $\rm T$ is illustrated by the orange arrow, whereas the hopping noise ${X}^\text{S}$ has almost no effect on the parafermionic qutrit.}
	\label{fig:qcm}
\end{figure}

To quantify the effect of errors, we exploit the behavior of the contextuality witness $\mathcal{M}$ given in (\ref{eq:resource}), which also quantifies the usefulness of the state in magic state distillation. For concreteness, we start with the state
\begin{equation}
	\rho=\ket{\psi}\!\!\bra{\psi} \,\, \text{with} \,\,
	\ket{\psi}=
	\begin{pmatrix}
		1/2  \\
		0  \\
		-\sqrt{3}/2 
	\end{pmatrix},
	\label{eqn:state}
\end{equation}
which is an almost optimal state for magic state distillation, encoded in the topological or non-topological qutrit. Then, we numerically analyze the response of the contextuality witness $\mathcal{M}$ with respect to joint hopping--phase noise in the parafermionic system and flip--phase noise in the non-topological system. These errors are described by the composed noise superoperators ${X}^\text{S}\circ{Z}^\text{S}$ and $\mathrm{T}\circ\Sigma$, respectively. Fig.\,\ref{fig:qcm}(a) shows that the presence of flip--phase noise quickly destroys the resource for magic state distillation in non-topological system. In contrast, the parafermionic qutrit is still free from any containment penetrating the degenerate subspace, as shown in Fig.\,\ref{fig:qcm}(b).

We now experimentally investigate the response of the parafermion edge zero mode qutrit state to the local noise in our photonic simulator and compare it with the effect of noise acting on a normal qutrit. 
We initially encode the state (\ref{eqn:state}) in our simulator. For simplicity, we restrict to hopping or flip errors as the full set of errors studied theoretically requires a larger Hilbert space than what can be supported by our simulator. This noise is simulated in our setup shown in Fig.~\ref{fig:setup} by removing the two $\mathcal{P}$ gates and keeping only the $\mathcal{R}$ gate. In particular, for the parafermionic qutrit we separate each of the three optical modes into two parts according to $p$, the probability of error to occur, and then project back into the computational basis with an imaginary-time evolution, which dissipates the modes outside the ground state subspace. For the non-topologically protected qutrit, we use the $\mathcal{R}$ gate to emulate the behavior of the superoperator $T$. The crucial difference in this case is that the modes in the excited subspace are no longer dissipated, resulting in the lift of the symmetry protection mechanism.

The experimental results for the contextuality witness $\mathcal{M}$ against noise are shown in Fig.\,\ref{fig:qcm}(c). From these data we see that for the parafermionic system, the witness does not decrease with respect to any noise strength, which is attributed to the topological protection. In particular, the parafermion encoded state remains an almost optimal resource state with $\mathcal{M}\geqslant0.580\pm0.013$. Indeed, when hopping noise acts on a parafermionic system, it only decreases the population of uncontaminated parafermions quasiparticles leaving the encoded states unaffected.  Consequently, the parafermion system is suitable for magic state distillation even in the presence of noise. 

The behavior of the normal qutrit under the presence of noise is rather different. From the experimental data, shown in Fig.\,\ref{fig:qcm}(c), we see that, for a normal qutrit, the value of ${\cal M}$ initially decreases for increasing error probability $p$ denoting that its quality as a resource for magic state distillation is diminished. At $p=\frac{2}{3}$, the inequality (\ref{eq:resource}) can no longer be violated and the noise has completely deprived the capability of the non-topological state for magic state distillation-based quantum computation. However, for sufficiently large error probabilities, $p>\frac{2}{3}$, the value for $\mathcal{M}$ increases again with $p$, because the maximum function in (\ref{eq:resource}) is obtained for another operator in $A^{\bf r}$. This behavior can be best understood from the dynamics of the states in the qubit Hilbert space. As is visualized in Fig.\,\ref{fig:qcm}(d), an exceeding amount of noise may eventually drive the state to the vicinity of another resource state of magic state distillation. We conclude that, unlike its non-topological counterpart, the qutrit encoded in the parafermion zero modes is able to protect contextuality and thus facilitates quantum computation with magic state distillation even for arbitrary levels of noise.

\subsection{Self-testing a topologically-protected qutrit in the face of local noise}

In the previous section, we used contextuality as an indicator for the applicability of magic state distillation. We now exploit contextually as a tool for \textit{self-testing} a quantum system. The term `self-testing' means that the observed correlations can only be explained, up to local isometries, by a specific state and set of measurements. In other words, a successful self-testing reveals a unique relationship between the state and the measurements. It has been shown that some NCHV inequalities from quantum contextuality are robust self-testings: if the choices of projective measurements are fixed, then the maximum violation of the NCHV inequality automatically singles out the form of the tested quantum state\,\cite{Bharti19}. The Klyachko-Can-Binicio\u{g}lu-Shumovsky (KCBS) inequality\,\cite{KCBS08} is such an example. Here we use the quantum violation of the KCBS inequality to self-test the encoded parafermionic qutrit in the face of local noise in order to further demonstrate its trait of noise-resilience.

The KCBS inequality is obtained by applying the graph-theoretic approach of contextuality\,\cite{CSW14} on a pentagon and it has an elementary form. Consider a set of measurements $B_i,\, i\in\{1,...,5\}$, where $B_i$ and $B_{i\text{\,mod\,5\,}+1}$ are orthogonal. We define $P_{\ket{\psi}}(B_i=1)$ as the probability of observing measurement outcome $B_i=1$ when the system is prepared in $\ket{\psi}$. Then, any NCHV theory assigning dichotomic values 1 for ``true'' and 0 for ``false'' regardless of context, satisfies
\begin{align}
	\mathcal{K} = \sum_{i=1}^5 P_{\ket{\psi}}(B_i=1) \overset{\text{NCHV}}{\leq} 2.
	\label{eq:kcbs}
\end{align}
However, a quantum qutrit system admits a maximum witness of contextuality $\mathcal{K}$ of $\sqrt{5}\approx{2.236}$. Hence, the KCBS inequality can be used as a robust self-testing: the maximum violation of the inequality can be achieved by only a single quantum state given the known, optimal set of projective measurements. Moreover, the difference between the experimentally obtained value and the maximum value of $\mathcal{K}$ lower bounds the trace distance between the optimal and actual input state.
\begin{figure}[tbp]
	\centering
	\includegraphics[width = .99\columnwidth]{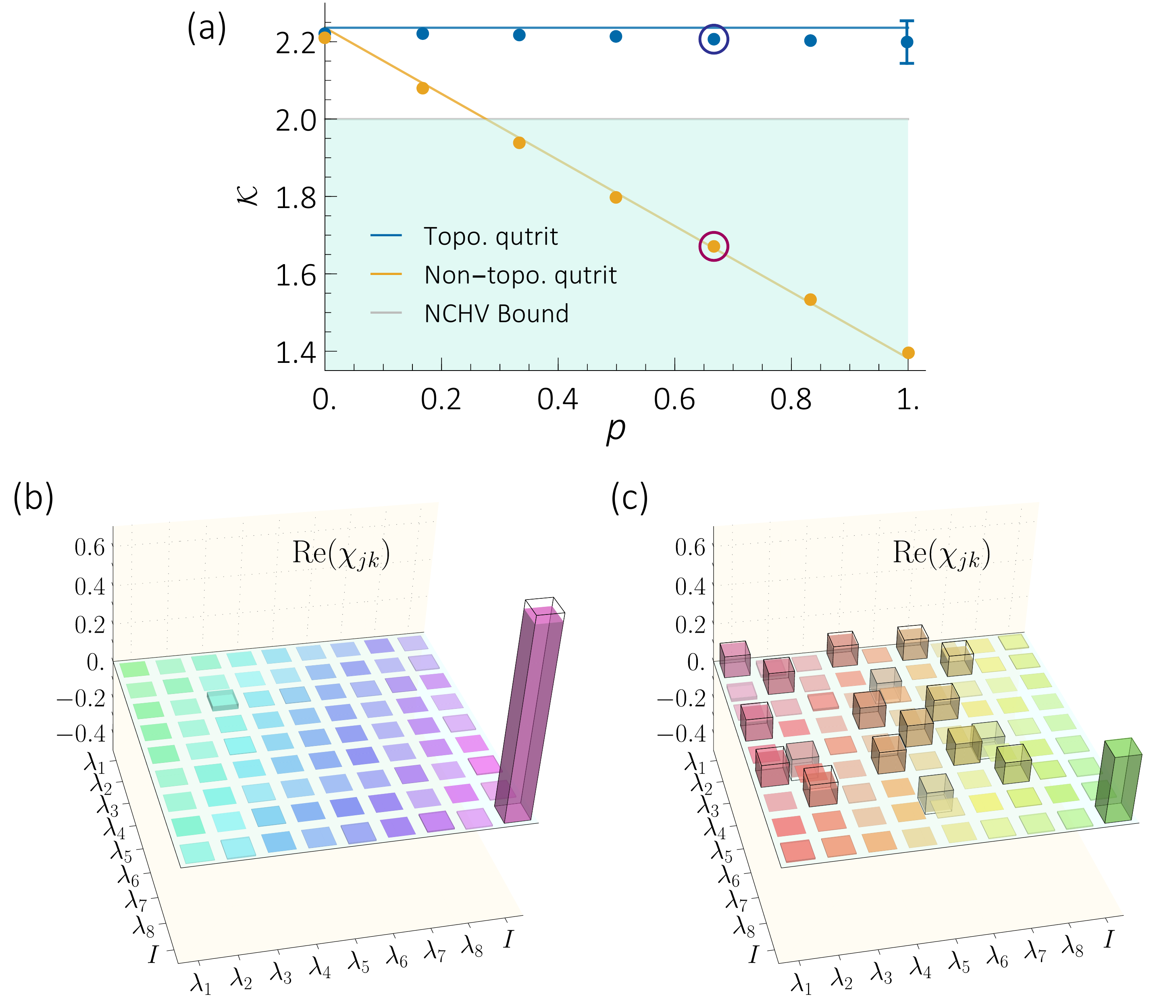}
	\caption{
		\textbf{Observation of topologically-protected quantum contextuality.}
		The local noise also reduces the violation of KCBS inequality quantified by ${\cal K}$, as is shown in (a). However, the violation of topologically-protected parafermions qutrit, denoted by the blue line and points in both cases, is almost unaffected.
		The real part of the process matrices, $\chi$, of the noise operators spanned on the Gell-Mann basis at the circled data points are displayed in (b) for the parafermions qutrit and (c) for non-topological qutrit. All elements in the imaginary part of the process matrices have absolute values of less than $1.25\times10^{-2}$. The edge and filling of the cuboids show the theoretical values and experimental results from quantum process tomography, respectively. The $1\sigma$ error bars deduced from a Poissonian counting statistics are too small to be not visible in the graph.}
	\label{fig:qck}
\end{figure}

We first derive theoretically the response of contextuality witness $\mathcal{K}$ to noise. In this subsection we again focus on the hopping and flip noises whose realizations are experimentally friendly. Without loss of generality, let the initial qutrit state be $\ket{\psi_2^\text{q}} = (0,\, 0,\, 1)^T$, and the five optimal measurement settings be $B_k=\ket{k}\bra{k}$, with
\begin{align}
	\bra{k}=\sqrt{1-\dfrac{1}{\sqrt{5}}} &\left(\cos \left(\dfrac{2k\pi}{5}\right),\, \sin \left(\dfrac{2k\pi}{5}\right),\, \dfrac{\sqrt{1+\sqrt{5}}}{2}\right), \nonumber\\
	k&=1, 2, ..., 5.
\end{align}
Then, under local disturbance $X^\text{S}$ and $\rm T$ on topological, $\ket{\psi_2^\text{S}}$, and non-topological, $\ket{\psi_2^\text{q}}$, qutrit, respectively, the value of ${\cal K}$ in (\ref{eq:kcbs}) becomes
\begin{align}
	\sum_{k=1}^5 \mathrm{Tr}\left[\ket{k}\!\!\bra{k}{X}^\text{S}\left(\ket{\psi_2^\text{S}}\!\!\bra{\psi_2^\text{S}}\right)\right] &= \sqrt{5}, \\
	\sum_{k=1}^5 \mathrm{Tr}\left[\ket{k}\!\!\bra{k}\mathrm{T}\left(\ket{\psi_2^\text{q}}\!\!\bra{\psi_2^\text{q}}\right)\right] &= \sqrt{5} -\frac{3\sqrt{5}-5}{2}p.
\end{align}
So for topologically encoded states, the violation of KCBS inequality is not affected by local noise, whereas for the normal qutrit states, the violation of (\ref{eq:kcbs}) is sensitive to the probability $p$ of local hopping errors.

Using the KCBS inequality, we investigate the resilience of contextuality granted by the topological qutrits. The experimental results are shown in Fig.\,\ref{fig:qck}(a). The value of $\mathcal{K}$ obtained from our measurements on the topological qutrit stays above 2.199 even under the effect of strong local noise. This value significantly violates the prediction of NCHV and confidently shows the existence of quantum contextuality. 
The proximity of the obtained $\mathcal{K}$ with its theoretical maximum value guarantees that the trace distance of our parafermions states before and after perturbation does not exceed the order of ${\cal O}(\sqrt{0.037})$\,\cite{Bharti19}. In comparison, when the flip noise with probability $p$ is applied on a normal qutrit, $\mathcal{K}$ quickly falls below $2$ as the error probability $p$ increases and the violation of the KCBS inequality fails. As a complement to our self-testing method, the process matrix $\chi_{jk}$ of noise operator acting on topological and trivial qutrit for $p=\frac{2}{3}$ is given in Fig.\,\ref{fig:qck}(b) and (c). For the topologically encoded qutrit the process matrix is the identity matrix with fidelity $96.0\%$ indicating the immunity of the parafermions states to local noise. On the other hand, the process matrix of the normal qutrit is significantly disturbed. 

The resilience of the parafermion qutrit is due to the increase of the system's energy caused by the application of noise and thus it is suppressed by the excitation gap. Equivalently, in the photonic parafermion quantum simulation, the effect of erroneous channels is dissipated by the imaginary time evolution. The observed noise-immunity clearly showcases the protection offered by topological order in the task of universal quantum computation. The observed reduction in the fidelity of the topologically encode qutrit is due to experimental imperfections. During our demonstration, such imperfections originate from phase mismatch of different photon paths or misalignment and imperfections of the photonic elements cause, on average, an overall reduction in fidelity of 4\%.

\section{Discussion and conclusions}

Topological systems are powerful candidates for future realizations of fault-tolerant quantum computation. A handicap present in both Majorana and parafermion-based systems for quantum computation is the operations created by braiding always belong to the Clifford group, which is not a universal set of quantum gates. To resolve this issue, we resort to magic state distillation as a pathway to acquiring universality. We have simulated two main elements of parafermions-based quantum computation under this architecture: the parafermions braiding that results in topological Clifford gates and the topologically-protected contextuality that enables the use of magic state distillation. While contextuality is a vital resource \textit{per se} for various quantum information tasks like one-way quantum computation\,\cite{Raussendorf13} and communication\,\cite{Saha19}, its interplay with topological system and noise resilience is not yet well understood. Our work may serve as a trailblazer for encoding contextuality in a real topological system and protecting it from environmental noise, thus benefiting the community of quantum information processing.

From a pragmatic perspective, simulation of a topological system with a realistic quantum simulator provides not only some novel insights on topological quantum computing, but also a testbed for investigating fault-tolerance in the encoded system. For example, quantum teleportation of Majorana zero modes with active error correction has been demonstrated on a superconducting quantum simulator\,\cite{xbzhu21}. Even though our photonic simulator is not exponentially scalable due to the use of a single photon for the encoding of the whole system, our method can be directly extended to several parafermions chains and further detect the contextuality of two qudits as proposed in\,\cite{Howard14}. Thus, it is plausible to expect the simulation of a few parafermions chains in the near future, where small topologically encoded algorithms can be performed in a fault-tolerant way\,\cite{ksun18}.

\begin{acknowledgments}
	The authors acknowledge insightful discussions with David Jennings and Penghao Zhu.
	This work was supported by
	National Key Research and Development Program of China (Grants No.\,2016YFA0302700, No.\,2017YFA0304100),
	the National Natural Science Foundation of China (Grants
	%Yong-Jian Han
	No.\,11874343,
	%%Xiao-Ye Xu
	%No.\,61805228,
	%Kai Sun
	No.\,61805227, No.\,61975195,
	%Chuan-Feng Li
	No.\,11774335, No.\,11821404,
	%Jin-Shi Xu
	No.\,61725504, and No.\,U19A2075),
	Key Research Program of Frontier Sciences, CAS (Grant No.\,QYZDY-SSW-SLH003),
	Science Foundation of the CAS (Grant No. ZDRW-XH-2019-1),
	the Fundamental Research Funds for the Central Universities (Grants No.\,WK2470000026, No.\,WK2030380017),  %
	Anhui Initiative in Quantum Information Technologies (Grants No.\,AHY020100, and No.\ AHY060300)
	and the EPSRC  (Grant No.\,EP/R020612/1).
	The data that support the findings of this study are available from the authors upon request.
	
\end{acknowledgments}

% -------------------- Supplementary Material --------------------

%\clearpage
%\newpage
%\onecolumngrid
\appendix

\setcounter{equation}{0}
\setcounter{figure}{0}
\renewcommand{\theequation}{S\arabic{equation}}
\renewcommand{\thefigure}{S\arabic{figure}}
\renewcommand{\thetable}{S\Roman{table}}

\section{Theoretical Details}

\subsection{Parafermions from the Fradkin--Kadanoff Transformation}

A pair of isolated Majorana zero modes can appear at the end of a chain of suitably coupled fermions\,\cite{Kitaev01}. A well-known mathematical equivalence, namely, the Jordan--Wigner transformation, can subsequently connect the fermionic chain to a chain of spin-1/2 particles.
The Jordan--Wigner transformation reads:
\begin{subequations}
	\begin{align}
		\gamma_{ka} &= \sigma_{z, k} \prod_{j<k} \sigma_{x, j}, \\
		\gamma_{kb} &= \sigma_{z, k} \prod_{j\leq k} \sigma_{x, j}.
	\end{align}
\end{subequations}
Where $\gamma_{km}$ denotes the \textit{creation and annihilation operator} of the Majorana fermion modes $m=a,b$ at the $k$-th site.
Perhaps the most notable feature of the Jordan--Wigner transformation is its non-locality. The constructed fermionic operator is defined with respect to an entire chain of spin operators located on its left hand sites. 
Due to the non-locality of the transformation, the local properties of the fermionic system are drastically different from those of the spin-1/2 system. Specifically, local noise on the spin system is no longer restricted to one site in the fermionic system. As a result the topological protection is removed.

Similarly to Majorana chains their generalization to $\mathbb{Z}_n$ parafermion chains (where $n=2$ corresponds to Majorana fermions) can be transformed to higher spin chain systems via the Fradkin--Kadanoff transformation.
First, the shift and clock operator, $\tau$ and $\sigma$, are defined as generalizations to the Pauli matrices $\sigma_x$ and $\sigma_z$ to higher dimensions:
\begin{align}
	\tau = \begin{pmatrix}
		0 & 0 & \dots & 0 & 1 \\
		1 & 0 & \dots & 0 & 0 \\
		0 & 1 & \dots & 0 & 0 \\
		\vdots & \vdots & \ddots & \vdots & \vdots \\
		0 & 0 & \dots & 1 & 0
	\end{pmatrix}, ~~~ \sigma &= \begin{pmatrix}
	1 & 0 & 0 & \dots & 0 \\
	0 & \omega & 0 & \dots & 0 \\
	0 & 0 & \omega^2 & \dots & 0 \\
	\vdots & \vdots & \vdots & \ddots & \vdots \\
	0 & 0 & 0 & \dots & \omega^{n-1}
\end{pmatrix}, \nonumber\\[12pt]
\text{with} ~ \omega &= e^{2\pi i/n}.
\end{align}
The two operators have the commutation relation of $\sigma \tau = \omega \tau \sigma$. In analogue to the Jordan--Wigner transformation, the pair of parafermion annihilation operators on the $k$-th site $\alpha_{ka}$ and $\alpha_{kb}$ is defined with the spin operators as:
\begin{align}
	\alpha_{ka} &= \sigma_k \prod_{j<k} \tau_j, \nonumber \\
	\alpha_{kb} &= \sigma_k \prod_{j\leq k} \tau_j.
	\label{eqs:FKT}
\end{align}
To demonstrate the effect of the transformation, substituting (\ref{eqs:FKT}) into the initial definitions of braiding Hamiltonians (main text formula (1)) yield the braiding Hamiltonians in the spin picture (main text formula (6)). The inverse transformation from parafermion picture to spin picture is:
\begin{align}
	\tau_k &= \alpha_{ka}^\dagger \alpha_{kb}, \nonumber \\
	\sigma_k =& \alpha_{ka} \prod_{j<k} \alpha_{jb}^\dagger \alpha_{ja}.
	\label{eqs:FKTi}
\end{align}
To demonstrate the effect of the inverse transformation consider the parity operator $Q =\displaystyle \prod_k \alpha_{ka}^\dagger \alpha_{kb}$. From the first subequation of (\ref{eqs:FKTi}) one finds that the form of parity operator in the spin picture is $\displaystyle \prod_k \tau_k$.

The Fradkin--Kadanoff transformation gives rise to exotic commutation relations for the parafermion operators; they neither commute nor anti-commute. Instead, $\alpha_{ka}\alpha_{la} = \omega^{\text{sgn} (l-k)} \alpha_{la}\alpha_{ka}$ and $\alpha_{kb}\alpha_{lb} = \omega^{\text{sgn} (l-k)} \alpha_{lb}\alpha_{kb}$ holds. Furthermore, the parafermion Hermitian conjugate relation $\alpha_{ka}^\dagger = \alpha_{ka}^{n-1}$ implies that unlike Majorana fermions, parafermions are not their own antiparticles.
Table.\,\ref{tab:comp} gives a comparison between Majorana fermions emerging as Majorana edge zero modes and parafermions as parafermion edge zero modes.

\begin{table*}[htbp]
	\centering
	\begin{tabular}{c|cc}
		\toprule
		& Majorana zero modes & Parafermion edge zero modes \\
		\midrule
		Symmetry & $\mathbb{Z}_2$ & $\mathbb{Z}_n$ \\
		Hamiltonian & $\mathcal{H}^\text{Mf} = if \displaystyle \sum_{k=1}^L \gamma_{ka} \gamma_{kb} + iJ \displaystyle \sum_{k=1}^{L-1} \gamma_{kb} \gamma_{(k+1)a}$, & $ \begin{aligned}
		\mathcal{H}^\text{Pf} = &-f e^{i\theta} \sum_{j=1}^L \alpha_{ja}^\dagger \alpha_{jb} \\ &-J e^{i\phi} \sum_{j=1}^{L-1} \alpha_{jb}^\dagger \alpha_{(j+1)a} + \text{h.c.}
		\end{aligned} $. \\
		Ground states & Twofold degenerate when $J>f>0$ & $n$-fold degenerate when zero modes exist \\
		Ground states crosstalk & Exponentially suppressed w.r.t. $L-1$ & Exponentially suppressed w.r.t. $L-1$. \\
		Involutory & Yes & No \\
		Parity & $(-1)^F = \displaystyle\prod_{k=1}^L (-i\gamma_{ka}\gamma_{kb})$ & $\omega^Q = \displaystyle\prod_{k=1}^L \alpha_{ka}^\dagger \alpha_{kb}$ \\
		Allowed states & Eigenstates of $(-1)^F$ & Eigenstates of $\omega^Q$ \\
		Homologous spin system & Ising model & Quantum Potts model \\
		\bottomrule
	\end{tabular}
	\caption{Comparison between Majorana fermions and parafermions.}
	\label{tab:comp}
\end{table*}

\subsection{Basis Rotation}

We use three sets of basis in the spin picture. Namely, $$\ket{0}_\sigma = \begin{pmatrix} 1 \\ 0 \\ 0 \end{pmatrix}, \ket{1}_\sigma = \begin{pmatrix} 0 \\ 1 \\ 0 \end{pmatrix}, \text{~and~} \ket{2}_\sigma = \begin{pmatrix} 0 \\ 0 \\ 1 \end{pmatrix}$$ are eigenstates of the clock operator $\sigma = \begin{pmatrix} 1 & 0 & 0 \\ 0 & \omega & 0 \\ 0 & 0 & \bar{\omega} \end{pmatrix}$; 
$$\ket{0}_\tau = \dfrac{1}{\sqrt{3}} \begin{pmatrix} 1 \\ 1 \\ 1 \end{pmatrix}, \ket{1}_\tau = \dfrac{1}{\sqrt{3}}\begin{pmatrix} 1 \\ \bar{\omega} \\ \omega \end{pmatrix}, \text{~and~} \ket{2}_\tau = \dfrac{1}{\sqrt{3}}\begin{pmatrix} 1 \\ \omega \\ \bar{\omega} \end{pmatrix}$$ are eigenstates of the shift operator $\tau= \begin{pmatrix} 0 & 0 & 1 \\ 1 & 0 & 0 \\ 0 & 1 & 0 \end{pmatrix}$; 
and $$\ket{0}_\chi = \dfrac{1}{\sqrt{3}} \begin{pmatrix} 1 \\ \omega \\ 1 \end{pmatrix}, \ket{1}_\chi = \dfrac{1}{\sqrt{3}}\begin{pmatrix} 1 \\ 1 \\ \omega \end{pmatrix}, \text{~and~} \ket{2}_\chi = \dfrac{1}{\sqrt{3}}\begin{pmatrix} 1 \\ \bar{\omega} \\ \bar{\omega} \end{pmatrix}$$ are the eigenstates of $\chi = \sigma\tau = \begin{pmatrix} 0 & 0 & 1 \\ \omega & 0 & 0 \\ 1 & \bar{\omega} & 0 \end{pmatrix}$. The eigenstates with index $k$ have eigenvalue $\omega^k$ with respect to their corresponding operator. 
Let us span a qutrit $\ket{\phi}$ on the three sets of bases $\ket{\phi}=A\ket{0}_\sigma + B\ket{1}_\sigma + C\ket{2}_\sigma = a\ket{0}_\tau + b\ket{1}_\tau + c\ket{2}_\tau = \alpha\ket{0}_\chi + \beta\ket{1}_\chi + \gamma\ket{2}_\chi$, then, the coefficients in each case are linked by:
\begin{align}
	\begin{pmatrix} a \\ b \\ c \end{pmatrix}  &= \frac{1}{\sqrt{3}}      \begin{pmatrix} 1 & 1 & 1 \\ 1 & \bar{\omega} & \omega \\ 1 & \omega & \bar{\omega} \end{pmatrix} \begin{pmatrix} A \\ B \\ C \end{pmatrix}, \\
	\begin{pmatrix} A \\ B \\ C \end{pmatrix} &= \frac{1}{\sqrt{3}} \begin{pmatrix} 1 & 1 & 1 \\ 1 & \omega & \bar{\omega} \\ 1 & \bar{\omega} & \omega \end{pmatrix} \begin{pmatrix} a \\ b \\ c \end{pmatrix}, \\
	\begin{pmatrix} \alpha \\ \beta \\ \gamma \end{pmatrix}  &= \frac{1}{\sqrt{3}} \begin{pmatrix} 1 & \bar{\omega} & 1 \\ 1 & 1 & \bar{\omega} \\ 1 & \omega & \omega \end{pmatrix} \begin{pmatrix} A \\ B \\ C \end{pmatrix}, \\
	\begin{pmatrix} A \\ B \\ C \end{pmatrix} &= \frac{1}{\sqrt{3}} \begin{pmatrix} 1 & 1 & 1 \\ \omega & 1 & \bar{\omega} \\ 1 & \omega & \bar{\omega} \end{pmatrix} \begin{pmatrix} \alpha \\ \beta \\ \gamma \end{pmatrix}.
\end{align}

To check the result of the basis rotation, we start from an arbitrary superposition of spin state composed of three ground states of $\mathcal{H}_0^\text{S}$, $\ket{\phi_0} = k_{00} \ket{0}_\sigma \ket{0}_\sigma \ket{0}_\sigma + k_{01}\ket{1}_\sigma \ket{1}_\sigma \ket{1}_\sigma + k_{02}\ket{2}_\sigma \ket{2}_\sigma \ket{2}_\sigma$.
In the first step of adiabatic evolution, the system is projected on $\mathcal{H}_1^\text{S}$'s ground state, the new ground state is $\ket{\phi_1} = \ket{0}_\tau(k_{10} \ket{0}_\sigma \ket{0}_\sigma + k_{11} \ket{1}_\sigma \ket{1}_\sigma + k_{12} \ket{2}_\sigma \ket{2}_\sigma)$. Use the relations of basis rotation, we conclude that
\begin{align}
	\begin{pmatrix} k_{10} \\ k_{11} \\ k_{12} \end{pmatrix}  &= \frac{1}{\sqrt{3}} \begin{pmatrix} 1 & 0 & 0 \\ 0 & 1 & 0 \\ 0 & 0 & 1 \end{pmatrix} \begin{pmatrix} k_{00} \\ k_{01} \\ k_{02} \end{pmatrix}.
\end{align}
The components that do not belong to the ground state of $\mathcal{H}_1^\text{S}$ have been discarded.

In the second step of adiabatic evolution, the system is projected on $\mathcal{H}_2^\text{S}$'s ground state, which is realized in two steps. Firstly, projecting on the first term of $\mathcal{H}_2^\text{S}$, which has no effect because the system is already in its ground state. 
Secondly, projecting on the second term of $\mathcal{H}_2^\text{S}$ yields the new ground state, which has the form of $\displaystyle \sum_{p, q=1}^{3} k_{2pq} \ket{p}_\sigma \ket{q}_\tau \ket{p+q\, \text{mod}\, 3}_\chi.$ By using the relations of basis rotation, we conclude that
\begin{align}
	\begin{pmatrix} k_{20q} \\ k_{21q} \\ k_{22q} \end{pmatrix}  &= \frac{1}{3\sqrt{3}} \begin{pmatrix} 1 & \bar{\omega} & 1 \\ 1 & 1 & \bar{\omega} \\ 1 & \omega & \omega \end{pmatrix} \begin{pmatrix} k_{10} \\ k_{11} \\ k_{12} \end{pmatrix}.
\end{align}
As the results for $q=0, 1\,\text{and}\,2$ are identical we can omit the subscript $q$ and rewrite the states as 
\begin{align*}
	\ket{\phi_2} =& k_{20} \ket{0}_\sigma (\ket{0}_\tau \ket{0}_\chi + \ket{1}_\tau \ket{1}_\chi + \ket{2}_\tau \ket{2}_\chi) \\
	&+ k_{21} \ket{1}_\sigma (\ket{2}_\tau \ket{0}_\chi + \ket{0}_\tau \ket{1}_\chi + \ket{1}_\tau \ket{2}_\chi) \\
	&+ k_{22} \ket{2}_\sigma (\ket{1}_\tau \ket{0}_\chi + \ket{2}_\tau \ket{1}_\chi + \ket{0}_\tau \ket{2}_\chi).
\end{align*}
In the third step of the adiabatic evolution, the system is projected on $\mathcal{H}_0^\text{S}$'s ground state, the new ground state is $\ket{\phi_3} = k_{30} \ket{0}_\sigma \ket{0}_\sigma \ket{0}_\sigma + k_{31}\ket{1}_\sigma \ket{1}_\sigma \ket{1}_\sigma + k_{32}\ket{2}_\sigma \ket{2}_\sigma \ket{2}_\sigma$. Projecting the $\ket{.}_\tau \ket{.}_\chi$ terms in the parentheses back to the $\sigma$-basis yields:
\begin{align}
	\begin{pmatrix} k_{30} \\ k_{31} \\ k_{32} \end{pmatrix}  &= 3\begin{pmatrix} 1 & 0 & 0 \\ 0 & 1 & 0 \\ 0 & 0 & \bar{\omega} \end{pmatrix} \begin{pmatrix} k_{20} \\ k_{21} \\ k_{22} \end{pmatrix}.
\end{align}
Experimentally, the three basis rotation are realized separately with an identity phase gate $\mathcal{P}_1 = \begin{pmatrix} 1 & 0 & 0 \\ 0 & 1 & 0 \\ 0 & 0 & 1 \end{pmatrix}$, a rotate gate $\mathcal{R}_2 = \frac{1}{\sqrt{3}} \begin{pmatrix} 1 & \bar{\omega} & 1 \\ 1 & 1 & \bar{\omega} \\ 1 & \omega & \omega \end{pmatrix}$ and another phase gate $\mathcal{P}_3 = \begin{pmatrix} 1 & 0 & 0 \\ 0 & 1 & 0 \\ 0 & 0 & \bar{\omega} \end{pmatrix}$, resulting in the matrix product of the final basis rotation. In the operational basis, the overall result of braiding matrix reads:
\begin{align}
	\tilde{\mathcal{B}} &= \mathcal{P}_3 \cdot \mathcal{R}_2 \cdot \mathcal{P}_1 = \frac{1}{\sqrt{3}} \begin{pmatrix} 1 & \bar{\omega} & 1 \\ 1 & 1 & \bar{\omega} \\ \bar{\omega} & 1 & 1 \end{pmatrix},
\end{align}
and an inverse Fourier transform $\mathcal{F}^\dagger = \displaystyle\frac{1}{\sqrt{3}} \begin{pmatrix} 1 & 1 & 1 \\ 1 & \bar{\omega} & \omega \\ 1 & \omega & \bar{\omega} \end{pmatrix}$ gives the result of braiding in the eigenbasis of $Q$:
\begin{align}
	\mathcal{B}^\text{S} &= \mathcal{F}^\dagger \cdot \tilde{\mathcal{B}} \cdot \mathcal{F} = e^{-{i\pi}/{6}} \begin{pmatrix} 1 & 0 & 0 \\ 0 & 1 & 0 \\ 0 & 0 & \omega \end{pmatrix}.
\end{align}

% \subsection{Effect of Braiding Operation on Quantum Contextuality}

\section{Experimental Details}

\subsection{Construction of Quantum Gates}

To implement the basis rotation, two sets of quantum gates, the phase gate $\mathcal{P}$, and the rotate gate $\mathcal{R}$, are employed. $\mathcal{P}$ adds different phases to different paths which together represent a photonic qutrit; $\mathcal{R}$ realizes a general transformation on a qutrit. 

Control of the relative phases in the $\mathcal{P}$ gate is realized by rotating the three independent half-wave plates between the two bulk quarter-wave plates acting on all of the three paths. Because the polarization states of photons in three paths are the same before entering the $\mathcal{P}$ gate, without loss of generality, we assume them to be horizontal.
The operations of the wave plates are best described by the Jones matrices. More precisely, the polarization states of a photon is denoted by $\ket{H} = \begin{pmatrix}1 \\ 0\end{pmatrix}$ for horizontal polarization and $\ket{V} = \begin{pmatrix}0 \\ 1\end{pmatrix}$ for vertical polarization. 
A half-wave plates is a birefriengent crystal, which adds an extra $\pi$ phase on photons whose polarization is perperdicular to its fast axis. The effect of half-wave plates with its fast axis at $\theta$ with respect to the horizontal axis can be represented by $hwp(\theta) = \begin{pmatrix}\cos 2\theta & \sin 2\theta \\ \sin 2\theta & -\cos 2\theta \end{pmatrix}$.
A quarter-wave plates is a birefriengent crystal, which adds an extra $\pi/2$ phase on photons whose polarization is perperdicular to its fast axis. The effect of quarter-wave plates with its fast axis at $\theta$ with respect to the horizontal axis can be represented by $qwp(\theta) = \begin{pmatrix}\cos^2 \theta + i\sin^2 \theta & (1-i) \sin \theta \cos \theta \\ (1-i) \sin \theta \cos \theta & \sin^2 \theta + i\cos^2 \theta \end{pmatrix}$. In the $\mathcal{P}$ gate, the quarter-wave plates are both aligned with their fast axes at $45^\circ$ with respect to the horizontal axis.

The effect of the $\mathcal{P}$ gate on one of the photon paths can be expressed as:
\begin{align*}
	\begin{pmatrix}1 \\ 0\end{pmatrix} &\to qwp\left(\frac{\pi}{4}\right) \cdot hwp(\theta) \cdot qwp\left(\frac{\pi}{4}\right) \cdot \begin{pmatrix}1 \\ 0\end{pmatrix} = \begin{pmatrix}e^{-2i\theta} \\ 0\end{pmatrix}.
\end{align*}
So a relative phase of $2\theta$ is added to this path when the half-wave plates is set at $\theta$ angle. Using the wave plates, individual phase can be precisely adjusted.

Control of the parameters in the $\mathcal{R}$ gate is realized by reshuffling the three optical modes and adjust the configuration of the new modes from the old ones. 
The three modes are first modulated by three independent half-wave plates set at different angles, so there polarization state is shifted from $\ket{H}$ to $hwp(\theta) \cdot \ket{H} = \begin{pmatrix} \cos 2\theta \\ \sin 2\theta \end{pmatrix}$. 
Next, a beam displacer (BD) horizontally separates the horizontally polarized component of each mode, so that three new optical modes are established. Repeating this process yields nine individual modes shaping a $3\times 3$ array, with their amplitude conveniently adjusted by rotating the orientation of the half-wave plates. At this point, tiltable glass plates are introduced as phase compensators to adjust relative phases between each mode. 
Finally, two BDs vertically merge the displaced beam modes twice, so the three modes on the left, middle and right sides of the array are combined. between and after the two combination, two half-wave plates orienting at $22.5^\circ$ and $\dfrac{1}{2} \arctan \left(\dfrac{1}{\sqrt{2}}\right) \approx 17.6^\circ$ adjust polarization of the photons and thus equalize the contribution of each mode into the final merged mode. The desired superposition is obtained with a polarizing beam splitter at the end of the $\mathcal{R}$ gate.

\begin{figure*}[htbp]
	\centering
	\includegraphics[width = .84 \textwidth]{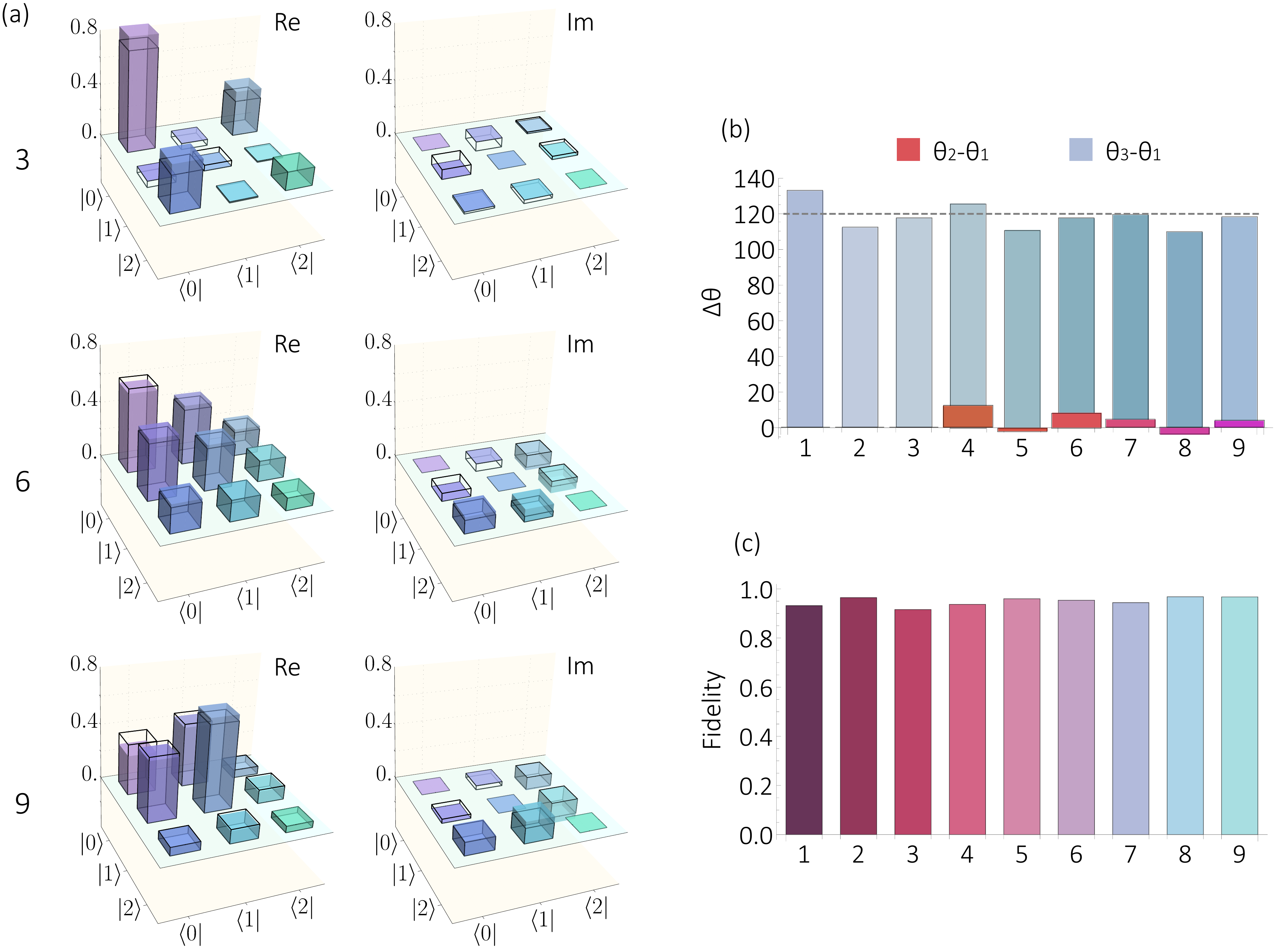}
	\caption{Additional experimental result on simulating braiding of parafermions. (a) Real and imaginary parts of sample density matrices for data points $\ket{\psi_3}$, $\ket{\psi_6}$, and $\ket{\psi_9}$ after braiding from quantum state tomography. The edge and the filling of the cuboids show the theoretical values and the experimental results from quantum process tomography, respectively. (b) Calculated phase change during the braiding evolution for each data point. The blue bars stand for the relative phase between $\ket{\psi_0^\text{S}}$ and $\ket{\psi_2^\text{S}}$, with a theoretical value of $2\pi/3$, and the red bars stand for the relative phase between $\ket{\psi_0^\text{S}}$ and $\ket{\psi_1^\text{S}}$, with a theoretical value of 0. (c) Calculated state fidelity between experimental result and theoretical prediction for each data point.}
	\label{figs:braidtomo}
\end{figure*}

\subsection{Quantum Process Tomography}

Braiding two parafermions ideally induces a unitary transformation that is diagonal on the parity operator's eigenbasis. Taking the experimental error and decoherence into consideration, the final form of the evolution is a completely positive trace-preserving map between initial and final states.
To completely characterize this map, we perform quantum process tomography and compare the acquired process matrix with its theoretical value to show the precision of applied operations.
The effect of a general completely positive trace-preserving map $\mathcal{E}$ acting on a density matrix $\rho$ can be expressed as $\rho \to \mathcal{E}(\rho) = \displaystyle \sum_{jk} \chi_{jk}\hat{E}_j \rho \hat{E}_k^\dagger$. Here, $\hat{E}$ is an orthonormal basis of the operators and $\chi$ is the process matrix. 
For a unitary evolution $U$, the theoretical value of $\chi$ can be linked by
$\chi_{jk} = c_j c_k^*$, where $\displaystyle\sum_j c_j \hat{E}_j = U$.

In our experiment, we reconstruct a completely positive trace-preserving map from a qutrit to a qutrit. The set of preparation and measurement basis are chosen to be 
%$\ket{\Psi_1}=\ket{0}, \ket{\Psi_2}=\ket{1}, \ket{\Psi_3}=\ket{2}, \ket{\psi_4}=\dfrac{1}{\sqrt{2}}(\ket{1}+\ket{2}), \ket{\Psi_5}=\dfrac{1}{\sqrt{2}}(\ket{0}+\ket{2}), \ket{\Psi_6}=\dfrac{1}{\sqrt{2}}(\ket{0}+\ket{1}), \ket{\Psi_7}=\dfrac{1}{\sqrt{2}}(\ket{1}-i\ket{2}), \ket{\Psi_8}=\dfrac{1}{\sqrt{2}}(\ket{0}-i\ket{2}),\, \text{and}\, \ket{\Psi_9}=\dfrac{1}{\sqrt{2}}(\ket{0}-i\ket{1})$
$\ket{0}, \ket{1}, \ket{2}, \dfrac{1}{\sqrt{2}}(\ket{1}+\ket{2}), \dfrac{1}{\sqrt{2}}(\ket{0}+\ket{2}), \dfrac{1}{\sqrt{2}}(\ket{0}+\ket{1}), \dfrac{1}{\sqrt{2}}(\ket{1}-i\ket{2}), \dfrac{1}{\sqrt{2}}(\ket{0}-i\ket{2}),\, \text{and}\, \dfrac{1}{\sqrt{2}}(\ket{0}-i\ket{1})$, where the set of eigenkets $\ket{0}, \ket{1}$ and $\ket{2}$ represents different path modes in the photonic simulator. 
The basis of Kraus operators are chosen to be the Gell-Mann matrices plus the identity matrix. Explicitly, they read:

\begin{figure*}[htbp]
	\centering
	\includegraphics[width = .84\textwidth]{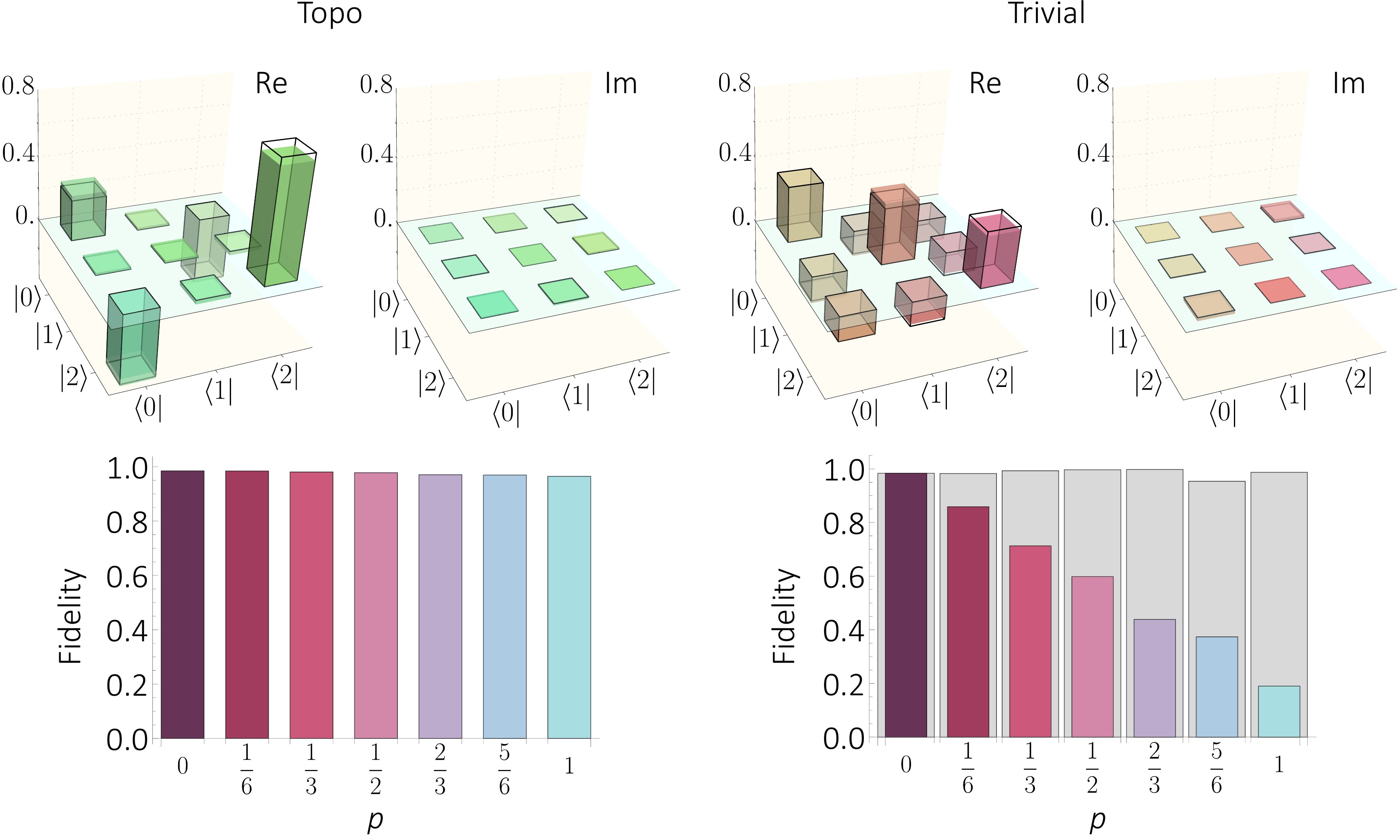}
	\caption{Additional experimental result on observation of contextuality in a parafermionic encoded system. The graph on the left side stands for result for parafermionic qutrits, and the graph on the right side stands for result for non-topological qutrits. 
		Top: real and imaginary parts of sample density matrices for the data points corresponding to $p=2/3$ hopping probability from quantum state tomography. The edge (filling) give the theoretical (experimental) values. 
		Bottom: calculated state fidelity between experimental result and theoretical prediction for each data point. In the graph on the right side, the gray and colored bars stand for fidelity with respect to the theoretical state when the effect of flip errors are taking and not taking into account in advance, respectively.}
	\label{figs:kcbstomo}
\end{figure*}

\begin{widetext}
	\begin{align}
		\hat{E}_1 = \lambda_1 = \begin{pmatrix} 0 & 1 & 0 \\ 1 & 0 & 0 \\ 0 & 0 & 0 \end{pmatrix}, ~~~\hat{E}_2 &= \lambda_2 = \begin{pmatrix} 0 & -i & 0 \\ i & 0 & 0 \\ 0 & 0 & 0 \end{pmatrix}, ~~~\hat{E}_3 = \lambda_3 = \begin{pmatrix} 1 & 0 & 0 \\ 0 & -1 & 0 \\ 0 & 0 & 0 \end{pmatrix}, ~~~ \nonumber \\ 
		\hat{E}_4 = \lambda_4 = \begin{pmatrix} 0 & 0 & 1 \\ 0 & 0 & 0 \\ 1 & 0 & 0 \end{pmatrix}, ~~~\hat{E}_5 &= \lambda_5 = \begin{pmatrix} 0 & 0 & -i \\ 0 & 0 & 0 \\ i & 0 & 0 \end{pmatrix}, ~~~\hat{E}_6 = \lambda_6 = \begin{pmatrix} 0 & 0 & 0 \\ 0 & 1 & 0 \\ 0 & 1 & 0 \end{pmatrix}, ~~~ \\
		\hat{E}_7 = \lambda_7 = \begin{pmatrix} 0 & 0 & 0 \\ 0 & 0 & -i \\ 0 & i & 0 \end{pmatrix}, ~~~\hat{E}_8 &= \lambda_8 = \frac{1}{\sqrt{3}} \begin{pmatrix} 1 & 0 & 0 \\ 0 & 1 & 0 \\ 0 & 0 & -2 \end{pmatrix}, ~~~\hat{E}_0 = \sqrt{\frac{2}{3}} I = \sqrt{\frac{2}{3}} \begin{pmatrix} 1 & 0 & 0 \\ 0 & 1 & 0 \\ 0 & 0 & 1 \end{pmatrix}. \nonumber
	\end{align}
\end{widetext}

To perform the quantum process tomography and obtain the process matrix $\chi$, we perform a least-square fit using a parameterized physical process matrix satisfying $\displaystyle \sum_{jk} \chi_{jk}\hat{E}_j \hat{E}_k^\dagger = 1$, and minimize the target function
\begin{align*}
	\sum_{m, n=1}^{9} \left[ p_{mn} - \chi_{jk} \sum_{j, k=0}^8 \braket{\tilde{\Phi}_n|\hat{E}_j|\tilde{\Psi}_m} \braket{\tilde{\Psi}_m|\hat{E}_k|\tilde{\Phi}_n} \right]^2
\end{align*}
with respect to the process matrix, where $p_{mn}$ is the detection probability when the input state is $\ket{\tilde{\Psi}_m}$ and the detection basis is $\bra{\tilde{\Phi}_n}$. In addition, the process matrix $\chi_{jk}^\text{S}$ in the eigenbasis of $Q$ can be expressed as $\chi_{jk}^\text{S} = \displaystyle\sum_{mn} \chi_{jk} \text{Tr} [(\mathcal{F}^\dagger \hat{E}_n \mathcal{F}) \hat{E}_i] \text{Tr} [(\mathcal{F}^\dagger \hat{E}_m \mathcal{F}) \hat{E}_j]$.

\subsection{More Results for Braiding}

In Fig.\,\ref{figs:braidtomo}, we present more experimental results on the photonic simulation of the PF braiding. Specifically, in subplot (a), three density matrices (corresponding to the quantum states $\ket{\psi_3}$, $\ket{\psi_6}$, and $\ket{\psi_9}$ in the main text) obtained after braiding are plotted together with their theoretical values. 
Subplot (b) shows the calculated phase change during the braiding evolution for each data point. The average increase of relative phase between $\ket{\psi_0^\text{S}}$ and $\ket{\psi_2^\text{S}}$ is $118.06^\circ$, and the average increase of relative phase between $\ket{\psi_0^\text{S}}$ and $\ket{\psi_1^\text{S}}$ is $3.80^\circ$. 
The calculated fidelities for each pair of theoretical and experimental states are plotted in (c). The values of fidelity range between 0.916 and 0.968.

\subsection{More Results for Contextuality and Noise Resilience}

In Fig.\,\ref{figs:kcbstomo}, we present more experimental results on the observation of contextuality in a parafermionic encoded system and the testing of the parafermionic system's robustness against local noise. In the top row, the density matrices obtained after applying a hopping error which happens with a probability of $2/3$ is plotted in the colored box, together with the theoretical values, plotted in the edge box. The graph with green bars stands for result for parafermionic qutrits, and the graph with red bars stands for result for non-topological qutrits. 
In the bottom row, we present the calculated state fidelity between experimental result and theoretical prediction at each flip probability. The average fidelity with respect to the undisturbed state for the topological qutrit is 0.98, while for the non-topologically encoded state the fidelity substantially decreases as the flip probability increase, and drops below 0.12 when the probability is 1. 
These results are plotted in the colored boxes. On the other hand, the fidelity for the trivial state with respect to the theoretical value when the hopping error has been taken into consideration is as high as 0.99. 

\begin{table}[htbp]
	\centering
	\vspace{20pt}
	\begin{tabular}{c|cc}
		\toprule
		$~i~$ & $P_{\ket{i}}(A_{i+1}=1)$ & $P_{\ket{i}}(A_{i-1}=1)$ \\ \midrule
		1 & $3.2\times 10^{-3}$ & $3.2\times 10^{-3}$ \\
		2 & $2.5\times 10^{-3}$ & $0.8\times 10^{-3}$ \\
		3 & $5.9\times 10^{-3}$ & $2.7\times 10^{-3}$ \\
		4 & $3.1\times 10^{-3}$ & $7.8\times 10^{-3}$ \\
		5 & $5.6\times 10^{-3}$ & $4.1\times 10^{-3}$ \\ \bottomrule
	\end{tabular}
	\caption{Experimentally measured orthogonality relationship of compatible measurements. By definition in (\ref{eq:kcbs}) in the main text, the conditions $\braket{i|i+1}=0$ and $\braket{i|i-1}=0$ are required by the compatiblity relationship.}
	\label{tab:qc-vert}
\end{table}

In a contextuality experiment, the orthogonality of measurement basis has to be checked in order to satisfy the requirement of compatibility and not open the signaling loophole. Table.\,\ref{tab:qc-vert} gives the measured detection probability when the preparation state is compatible with the target detection basis, so the detection probabilities on these bases are ideally 0.
These measured detection probabilities are of the order of $10^{-3}$ in accordance with the requirement of compatibility.

\bibliography{references}

\end{document}